# The Hidden Structure of the Factor Zoo: When Asset Pricing Meets Kolmogorov–Arnold

Peng Liu[*] Mengyu Wang[†] Youyou Xu[‡] Hong Zhang[§]

This Draft: Nov 16, 2026

**Abstract**

How many independent sources of priced risk are embedded in firm characteristics, and why do existing estimates differ so widely? We address these questions using the Kolmogorov–Arnold representation theorem to develop a two-layer Kolmogorov–Arnold Asset Pricing Model (KAPM): an inner layer that aggregates firm characteristics into synthetic factors and an outer layer that maps these factors into expected returns. We estimate KAPM using the Kolmogorov–Arnold Network (KAN), a state-of-the-art, interpretable machine-learning implementation of the theorem. The selected architecture contains eight synthetic factors, which deliver strong out-of-sample return predictability. A Bayesian analysis of these factors further favors a more parsimonious three-factor model that can explain the cross-section of stock returns, a task that traditional factors and individual anomalies fail to accomplish. Further analysis yields two important asset-pricing implications. First, layer-wise ablation tests compare several classes of asset-pricing models nested in KAPM and show that nonlinear effects of individual characteristics, even without interactions among them, suffice to achieve most of KAPM's out-of-sample portfolio performance. Second, KAPM factors draw on information from a broad range of anomaly themes, suggesting that its parsimonious factor structure does not imply that only a few anomalies matter. Overall, our results suggest that the parsimonious nonlinear structure of KAPM may suffice to capture the pricing-relevant information embedded in the factor zoo to explain the cross-section of asset returns.

[*] Liu: Lee Kong Chian School of Business, Singapore Management University. Email: liupeng@smu.edu.sg.
[†] Wang: Lee Kong Chian School of Business, Singapore Management University. Email: mywang@smu.edu.sg.
[‡] Xu: Lee Kong Chian School of Business, Singapore Management University. Email: youyou.xu.2025@pbs.smu.edu.sg.
[§] Zhang: Lee Kong Chian School of Business, Singapore Management University. Email: hongzhang@smu.edu.sg.

## I. Introduction

Empirical asset pricing studies have produced hundreds of return predictors (see, among others, McLean and Pontiff 2016, Hou, Xue, and Zhang 2020, and Jensen, Kelly, and Pedersen 2023 for cross-country validation). This proliferation creates what Cochrane (2011) famously termed the "factor zoo", and raises a fundamental question: how many independent sources of priced risk are actually embedded in firm characteristics? Existing answers vary widely. Multiple-testing approaches impose more stringent statistical criteria to distinguish genuine return predictors from false discoveries (Harvey, Liu, and Zhu 2016; Feng, Giglio, and Xiu 2020). Economically motivated factor models seek to summarize the cross-section of returns with a small set of factors derived from economic principles (Fama and French 2015; Hou, Xue, and Zhang 2015; Hou, Mo, Xue, and Zhang 2021). Statistical factor-extraction methods instead use tools such as principal components and instrumented principal components to recover a small number of latent factors from a large set of characteristics (Chamberlain and Rothschild 1983; Kelly, Pruitt, and Su 2019; Kozak, Nagel, and Santosh 2020). Beyond these linear approaches, nonlinear characteristic-based models allow characteristics to enter the expected-return function flexibly (Connor, Hagmann, and Linton 2012; Freyberger, Neuhierl, and Weber 2020), while nonlinear pricing-kernel models relax linearity in the mapping from systematic risks to asset prices (Bansal and Viswanathan 1993; Rosenberg and Engle 2002). Because these approaches impose different restrictions on the structure of the mapping from characteristics to expected returns, they need not imply the same dimensionality. The central challenge is therefore not only to measure dimensionality, but also to understand the underlying asset-pricing structure that determines it.

Our paper aims to address this challenge by proposing a broad asset pricing framework based on the Kolmogorov–Arnold representation theorem (Kolmogorov 1957; hereafter, KA), which we term the *Kolmogorov–Arnold Asset Pricing Model* (KAPM). The KA theorem states that, under general conditions, any continuous multivariate function can be represented as a two-layer composition of univariate functions. Applying this mathematical theorem to asset pricing gives rise to a general two-layer framework for determining expected asset returns. The inner layer aggregates firm characteristics into a set of synthetic factors through additive combinations of univariate characteristic transformations. The outer layer then maps these synthetic factors into expected returns through a second set of univariate functions. Intuitively, the two layers resemble the nonlinear characteristic-based models and nonlinear pricing-kernel models, respectively. The universal representation property of the KA theorem, however, unifies these two classes of models by suggesting that, taken together, the two-layer framework can represent an arbitrary continuous characteristic-to-return asset pricing function.

Because the KAPM allows for nonlinear functions in both layers, traditional linear statistical tools are not well suited to estimating the system. Hence, we employ a state-of-the-art machine-learning algorithm, the Kolmogorov–Arnold Network (Liu et al. 2025; hereafter, KAN), to estimate KAPM. More explicitly, we use as inputs the 153-anomaly dataset of Jensen, Kelly, and Pedersen (2023), augmented with interactions between these anomalies and four macroeconomic state variables obtained from Goyal, Welch, and Zafirov (2024). The interaction approach follows Gu, Kelly, and Xiu (2020; hereafter GKX). Also following GKX's expanding-window procedure, we initially estimate KAPM using 30 years of data (1957–1986) and use it to predict stock returns. We then recursively re-estimate KAPM each year (using all available data from 1957 through the preceding year) to generate out-of-sample return forecasts from 1987 through 2023.[1] To avoid the well-known small-stock bias in machine-learning asset-pricing tests, our main analysis excludes stocks below the NYSE 20th-percentile size breakpoint.

The model's most important hyperparameters include the number of synthetic factors, the spline order, the network depth, and the degree of regularization. Since several of these hyperparameters have direct implications for the structure and dimensionality of KAPM, we select them through a rigorous architecture-selection procedure as detailed in the Internet Appendix, including an in-sample Sharpe-ratio horse race and Bayesian factor-inclusion tests. The selected architecture contains eight synthetic factors (hereafter, the eight-factor KAPM or KAPM8), has a network depth of one, and uses cubic splines. This implies that a relatively shallow network with a small number of latent factors suffices to capture the in-sample characteristic-return mapping.

We next examine the out-of-sample performance of KAPM8. We first conduct portfolio analysis. Each month, we sort stocks into deciles based on their one-month-ahead KAPM return forecasts and form a high-minus-low portfolio that buys the highest-predicted-return decile and sells the lowest. The portfolios are rebalanced monthly. Over the 1987–2023 out-of-sample period, the value-weighted and equal-weighted KAPM8 long–short portfolios achieve annualized Sharpe ratios of 1.38 and 1.68, comparable to the best neural-network specification of GKX. The value-weighted and equal-weighted portfolios also deliver annualized returns of 20.2% and 22.1%, respectively, far exceeding the corresponding returns of 2.39% and 3.03% for the average anomaly portfolio over the same period. We then measure risk-adjusted performance using a six-factor model (hereafter FF6) based on the Fama–French (2015) five factors augmented with the momentum factor (Carhart 1997). The two portfolios earn monthly FF6 alphas of 1.74% and 1.78%, respectively, both economically large and statistically significant.

[1] Within each testing year, the estimated model parameters are held fixed. The latest firm characteristics and macroeconomic state variables are updated monthly to generate one-month-ahead return forecasts during the year. The model is then re-estimated at the end of each year using the expanded training window.

After portfolio analysis, we apply cross-sectional asset-pricing tests to KAPM factors. We start with KAPM8. Following the tradition of cross-sectional tests using time-series average returns (e.g., Black, Jensen, and Scholes 1972; Adrian, Etula, and Muir 2014; He, Kelly, and Manela 2017), we examine whether KAPM8-implied expected returns explain the cross-section of individual stocks' *average* realized excess returns over the 1987–2023 out-of-sample period. We also report Fama–MacBeth (1973) t-statistics based on monthly cross-sectional regressions (e.g., Adrian, Etula, and Muir 2014; Borri et al. 2024). If the KAPM8 prediction were perfect, we would expect a regression coefficient of one. Empirically, using the KAPM8 prediction, we obtain a positive and highly significant coefficient close to one in magnitude (1.136, t=4.47), indicating that KAPM8-implied expected returns closely align with average realized returns in the cross-section. Furthermore, adding conventional factor exposures (CAPM, FF3, FF5, and FF6) does not absorb this pricing power. Indeed, the economic magnitude and statistical significance of the coefficient remain largely unchanged across all specifications. Adding the KAPM8 signal to the six-factor model also raises the cross-sectional adjusted $R^2$ from about 0.5% to 8.3%.

One caveat is that the intercepts remain significant across these specifications, suggesting that neither the six-factor model nor the KAPM8 prediction can fully explain the cross-section of stock returns. Even when the KAPM8 prediction and the six-factor model are used jointly, the cross-sectional regression exhibits a negative and significant intercept of −0.273 (t = −2.32).

Despite the strong performance of KAPM8, the common concern of in-sample overfitting raises an important question: whether all eight synthetic factors are necessary to capture its out-of-sample pricing power. Indeed, the out-of-sample Sharpe ratio of high-minus-low portfolios formed on KAPM-predicted returns initially increases as more synthetic factors are included. It then peaks with the three leading factors and no longer increases as additional factors are added. This pattern suggests that the remaining factors may not contain independent information for explaining out-of-sample (OOS) asset returns. This observation motivates us to identify a more parsimonious factor structure by applying the Bayesian model-selection framework of Barillas and Shanken (2018), as developed further by Chib, Zeng, and Zhao (2020; see also Bryzgalova, Huang, and Julliard 2023). The full-sample Bayesian analysis favors an even more parsimonious specification consisting of the three leading synthetic factors (hereafter, the three-factor KAPM or KAPM3).

KAPM3 delivers even better risk-adjusted performance, with monthly FF6 alphas of 1.75% and 1.82% for the value-weighted and equal-weighted portfolios, respectively. Its annualized Sharpe ratios are 1.43 and 1.78, compared with 1.38 and 1.68 for KAPM8. Figure 1 compares the cumulative out-of-sample returns of the average anomaly portfolio, KAPM8, and KAPM3, based on their equally weighted long–short portfolios over 1987–2023. Both KAPM portfolios accumulate far larger returns than the anomaly benchmark. Between the two KAPM portfolios, KAPM3 outperforms. The leading three factors of KAPM3 thus carry most of the model's economic value, while the remaining five provide little incremental out-of-sample performance.

Interestingly, KAPM3 largely eliminates the pricing intercept that KAPM8 and traditional asset pricing models fail to explain. We first demonstrate this effect by examining the cross-sectional pricing power of the three leading factors as separate regressors. These three factors exhibit cross-sectional explanatory power beyond conventional factor exposures. More importantly, when the three factors enter jointly, the intercept becomes economically small and statistically insignificant. For example, the FF6-adjusted intercept is 0.141% ($t = 0.22$). Using the aggregate return prediction of KAPM3 yields a similar result, with an FF6-adjusted intercept of 0.048% ($t = 0.19$). These results suggest that KAPM3 captures important components of average stock returns that conventional factors leave unexplained.

We further compare KAPM3 with a broader set of asset-pricing factors. These include economic factors based on the q-factor model (Hou, Xue, and Zhang 2015), traded liquidity (Pástor and Stambaugh 2003), quality-minus-junk (Asness, Frazzini, and Pedersen 2019), betting-against-beta (Frazzini and Pedersen 2014), and mispricing factors (Stambaugh and Yuan 2017), as well as statistical factors based on principal component analysis (PCA; Kozak, Nagel, and Santosh 2018, 2020), risk-premium PCA (RP-PCA; Lettau and Pelger 2020), kernel PCA (K-PCA; Schölkopf, Smola, and Müller 1998), and instrumented principal component analysis (IPCA; Kelly, Pruitt, and Su 2019). For economic factors, we adopt a three-step analysis: 1) augment the FF3 factors with each alternative factor, 2) estimate stock-level exposures to these benchmark factors, and 3) run cross-sectional regressions of average returns, first using the benchmark factors alone and then adding the three KAPM3 factors. For the statistical models, we replace the first step with the three leading factors extracted by each method and then follow the same second and third steps. When used alone, each benchmark specification leaves a significant cross-sectional intercept. In contrast, adding the KAPM3 factors renders the intercept statistically insignificant in every case. All KAPM3 factors exhibit significant cross-sectional explanatory power at least at the 10% level.

Likewise, we also ask whether KAPM3 prices the cross-section of asset returns better than the individual anomalies from which it is constructed. When used alone, the individual anomaly exposure leaves a significant cross-sectional intercept unexplained (at the 10% level) in about three-quarters of the cases. Moreover, the anomaly exposures have little cross-sectional explanatory power on average, with an average absolute t-statistic of only 0.61.[2] In contrast, adding the three KAPM3 factors renders the intercept statistically insignificant in every case, while all three KAPM3 factors exhibit significant cross-sectional explanatory power at least at the 10% level.

Taken together, our asset pricing tests suggest that KAPM3 explains the cross-section of average asset returns above and beyond commonly used factors and individual anomalies. These results point to a remarkably low-dimensional asset-pricing structure: although KAPM8 retains eight synthetic factors, three leading factors appear sufficient to explain the cross-section of average stock returns in our tests.

After confirming the asset-pricing power of KAPM, we now explore two important implications for understanding the factor zoo. First, since KAPM differs from linear factor models primarily in its nonlinear use of information, we want to shed light on the economic sources of nonlinearity. As discussed above, the two layers of KAPM intuitively resemble two distinct forms of nonlinearity studied in the asset-pricing literature. The inner layer resembles nonlinear characteristic-based models by allowing characteristics to enter the construction of synthetic factors nonlinearly, whereas the outer layer resembles nonlinear pricing-kernel models by allowing the synthetic factors to be priced nonlinearly. We therefore conduct a layer-wise ablation analysis that selectively removes each source of nonlinearity.

[2] The anomaly exposures have an average absolute slope t-statistic of 0.61 and a median of 0.52. The 90th percentile is 1.17 and the 99th percentile is 1.58, so that almost none of the individual anomalies exhibit statistically significant cross-sectional explanatory power at the 10% level.

Our comparison starts from the fully linear benchmark (denoted as L/L), which we proxy by the high-minus-low portfolios of either the average anomalies or the three leading PCA factors. PCA provides a particularly appealing benchmark because it can be intuitively viewed as a two-layer linear structure: the inner layer aggregates information from the factor zoo into a small number of factors through linear transformations, while the outer layer links these factors to expected returns through linear factor exposures. We then allow for nonlinearity in each of the two layers separately, while restricting the other layer of the estimated KAPM to be linear.

This exercise allows us to assess not only the gains from nonlinearity, but also which form of nonlinearity accounts for KAPM's pricing performance. To illustrate the intuition, we focus on the NL/L specification, which allows nonlinear characteristic aggregation while restricting the outer-layer pricing function to be linear. On the one hand, the nonlinear inner layer nests nonparametric characteristic-based models such as Freyberger, Neuhierl, and Weber (2020), allowing individual characteristics to affect expected returns through nonlinear transformations. An important feature of this layer is the absence of interactions across different characteristics.[3] On the other hand, since the outer layer remains linear, it does not introduce additional cross-characteristic interactions. Overall, therefore, the predictive gains of NL/L relative to the linear benchmark can be attributed to additive nonlinear effects of individual characteristics, rather than interactions among them.

Empirically, the NL/L ablation increases the VW Sharpe ratio from 0.52 (PCA) to 1.19 (NL/L KAPM8) and 1.26 (NL/L KAPM3). The corresponding EW Sharpe ratio increases from 0.69 to 1.47 and 1.54, respectively. Hence, allowing for nonlinearity in the inner layer more than doubles the Sharpe ratios relative to the linear PCA benchmark. Compared to the Sharpe ratios of 1.38 and 1.68 for the fully nonlinear KAPM8 (NL/NL), the NL/L specification captures approximately 78% of the VW improvement and 79% of the EW improvement over the PCA benchmark. The corresponding figures for KAPM3 are 81% and 78%, respectively. The NL/L specifications also generate substantial risk-adjusted returns, with monthly FF6 alphas of 1.36% (VW) and 1.52% (EW) for KAPM8, and 1.45% and 1.58% for KAPM3, respectively. These alphas not only substantially exceed those of the PCA benchmark (0.76% VW and 0.89% EW), but also capture most of the alphas achieved by the fully nonlinear KAPM models (NL/NL), as reported above.

[3] We also use the characteristics-macroeconomic interactions as return predictors. Since these interactions can be interpreted as characteristics affecting the firm's exposure to macroeconomic conditions (e.g., Kelly, Pruitt, and Su 2019), we interpret their effects as contributions of individual characteristics.

Furthermore, while the NL/L specification restricts the return impact of synthetic factors to be linear, our cross-sectional tests show that the three NL/L KAPM3 factors jointly leave no statistically significant intercept, even after controlling for conventional factor exposures. Taken together, these results suggest that additive nonlinear effects of individual characteristics, without requiring cross-characteristic interactions, suffice to capture most of KAPM's out-of-sample portfolio performance and cross-sectional pricing power.[4]

The second implication concerns the economic interpretation of KAPM's parsimonious factor structure. The existence of a low-dimensional factor model is consistent with two different interpretations. First, only a small subset of anomalies may contain pricing-relevant information, while the remaining anomalies are largely redundant (e.g., Fama and French 2015; Hou, Xue, and Zhang 2015; Hou, Mo, Xue, and Zhang 2021). Alternatively, pricing-relevant information may be dispersed across a broad range of anomalies but can be aggregated into a small number of synthetic factors (e.g., Kozak, Nagel, and Santosh 2018, 2020).

To distinguish between these interpretations, we examine how KAPM3's low-dimensional factor structure emerges from the factor zoo. An important advantage of KAN is that the transformations in both layers can be approximated by polynomial functions, allowing us to trace each synthetic factor back to the underlying anomalies and their economic themes. Using the anomaly-theme classification of Jensen, Kelly, and Pedersen (2023), we examine how different themes contribute to the synthetic factors. For the inner-layer mapping, we observe that the first factor places somewhat greater weight on momentum and short-term reversal, the second on momentum and low risk, and the third on low risk and short-term reversal. Nevertheless, no single theme dominates any of the three factors.

[4] This observation is consistent with the recent findings of Lu, Spiegel, and Zhang (2025), who show that much of neural-network performance can be explained by a dynamically managed arbitrage portfolio loading on squared anomaly alphas. In contrast, the L/NL case lacks such a clean interpretation. Here, each synthetic factor is a linear combination of firm characteristics (L in the inner layer), while the mapping from synthetic factors to returns is nonlinear (NL in the outer layer). In this case, even a simple squared synthetic factor combines the higher-order contribution of each individual characteristic with pairwise interaction terms among them, so the two cannot be separately identified. Nonetheless, the improvement from L/L to L/NL captures the benefit of nonlinear pricing kernels or stochastic discount factors documented in the literature (Bansal and Viswanathan 1993; Aït-Sahalia and Lo 2000; Rosenberg and Engle 2002).

These results support the second interpretation: KAPM's parsimonious factor structure reflects the nonlinear aggregation of pricing-relevant information dispersed across many anomalies, rather than the selection of a few dominant predictors. Together with the layer-wise ablation analysis, these findings suggest that a proper estimation of the hidden structure of the factor zoo may help reveal how information from the factor zoo is aggregated into a few factors and how these factors help explain the cross-section of asset returns.

Our paper speaks to several strands of literature. First, we contribute to the literature on the factor zoo by proposing a general asset-pricing framework that nests existing approaches within a two-layer representation. When both layers are linear, KAPM encompasses conventional linear factor structures (Fama and French 2015; Hou, Xue, and Zhang 2015; Hou, Mo, Xue, and Zhang 2021; Kelly, Pruitt, and Su 2019; Kozak, Nagel, and Santosh 2018, 2020). Allowing nonlinearity in the inner and outer layers resembles nonlinear characteristic-based models (Connor, Hagmann, and Linton 2012; Freyberger, Neuhierl, and Weber 2020) and nonlinear pricing-kernel models (Bansal and Viswanathan 1993; Aït-Sahalia and Lo 2000; Rosenberg and Engle 2002), respectively. This framework allows us both to identify a remarkably low-dimensional pricing structure and to demonstrate how information dispersed across the factor zoo can be nonlinearly compressed into a small number of synthetic factors to explain the cross-section of asset returns.

Our implementation of KAPM also connects to the rapidly expanding literature on machine learning in empirical asset pricing. Existing studies use neural-network and tree-based nonlinear machine-learning tools to improve return prediction (e.g., Gu, Kelly, and Xiu 2020, 2021; Bryzgalova, Pelger, and Zhu 2025; Cong, Feng, He, and He 2025) and explore the importance of model complexity and richer factor representations (Kelly, Malamud, and Zhou 2024; Didisheim, Ke, Kelly, and Malamud 2023, 2024).[5] Our findings suggest that a highly flexible nonlinear model can nevertheless reveal a low-dimensional factor structure. Recent work applies KAN to conditional factor models (Wang and Singh 2024). We are the first, to the best of our knowledge, to use the state-of-the-art machine-learning algorithm KAN (Liu et al. 2025) to estimate a two-layer Kolmogorov–Arnold Asset Pricing Model (KAPM) and uncover the nonlinear factor structure underlying the factor zoo. In the machine-learning literature, conventional neural networks and KAN provide two parallel machine-learning architectures built on different mathematical foundations: the Universal Approximation Theorem (Cybenko 1989; Hornik, Stinchcombe, and White 1989; Hornik 1991) and the Kolmogorov–Arnold Representation Theorem (Kolmogorov 1957; Liu et al. 2025), respectively. While the two approaches exhibit similar power in predicting stock returns, KAPM estimated using KAN has the unique advantage of revealing the hidden structure of the factor zoo and providing interpretable estimation results that alleviate the "black-box" concern associated with machine learning.

[5] Kelly and Xiu (2023) provide a comprehensive recent survey. Also see recent applications of nonlinear machine learning to cross-sectional asset-return prediction in Chen, Pelger, and Zhu (2024), Murray, Xia, and Xiao (2024), Li, Rossi, Yan, and Zheng (2025), and Bell, Kakhbod, Lettau, and Nazemi (2026). In particular, Bell, Kakhbod, Lettau, and Nazemi (2026) use an Explainable Boosting Machine (EBM) to directly estimate predictor nonlinearities and interactions for corporate bond returns. KAPM provides a complementary approach that embeds nonlinear functional forms within a parsimonious factor structure, allowing us to characterize nonlinearities while focusing on a low-dimensional asset-pricing framework.

Our paper is also closely related to Borri, Chetverikov, Liu, and Tsyvinski (2024), who apply the Kolmogorov–Arnold representation theorem to create a single nonlinear asset pricing factor. We differ by retaining the original two-layer representation of the KA theorem and using KAN to estimate the underlying structure. This two-layer structure has the advantage of nesting existing asset pricing frameworks, allowing our analysis to uncover the dimensionality of the factor zoo, distinguish the sources of nonlinearity, and trace the parsimonious KAPM factors back to the underlying anomalies.

The remainder of the paper is organized as follows. Section II develops the KAPM framework and its estimation via KAN. Section III describes the data and empirical design. Section IV evaluates the out-of-sample performance of KAPM8. Section V examines the dimensionality of KAPM and the asset-pricing performance of KAPM3. Section VI investigates the sources of nonlinearity and the economic interpretation of the synthetic factors. Section VII concludes.

## II. The Kolmogorov–Arnold Asset Pricing Model

We now develop the Kolmogorov–Arnold Asset Pricing Model (KAPM) and describe its estimation using KAN.

### A. A Two-Layer Representation of Expected Returns

We begin with the standard stochastic discount factor (SDF) representation of asset returns. Let $r_{i,t+1}$ denote the excess return of asset $i$ from period $t$ to $t+1$. The no-arbitrage condition requires

$$E_t[m_{t+1} r_{i,t+1}] = 0, \tag{1}$$

where $m_{t+1}$ denotes the stochastic discount factor. Applying the covariance decomposition, the condition can be rewritten as

$$E_t[r_{i,t+1}] = -\frac{1}{E_t[m_{t+1}]} Cov_t(m_{t+1}, r_{i,t+1}). \tag{2}$$

Assume that a set of $K$ common factors $\mathbf{f}_{t+1} \in \mathbb{R}^K$ drives asset returns and is associated with time-varying factor risk premia $\boldsymbol{\lambda}_t \in \mathbb{R}^K$. Let $\mathbf{Z}_{i,t} \in \mathbb{R}^L$ denote the vector of observable firm characteristics that determine the factor exposure of asset $i$. The vector $\mathbf{Z}_{i,t}$ may include interactions between firm characteristics and aggregate state variables, which we treat as time-varying firm characteristics.

Given the information available at time $t$, firm characteristics $\mathbf{Z}_{i,t}$ are known. We assume that these characteristics and the time-varying factor risk premia sufficiently characterize the conditional covariance between the stochastic discount factor and asset returns. Hence, we write

$$Cov_t(m_{t+1}, r_{i,t+1}) = C(\mathbf{Z}_{i,t}, \boldsymbol{\lambda}_t), \tag{3}$$

where $C(\cdot)$ is an unknown continuous function. Moreover, pricing the risk-free asset implies

$$E_t[m_{t+1}] = 1/R_{f,t}, \tag{4}$$

where $R_{f,t}$ denotes the gross risk-free rate over the next period, which is known at time $t$. Combining equations (2)–(4), the conditional expected return is

$$E_t\big[r_{i,t+1}\big] = -R_{f,t}\, C\big(\mathbf{Z}_{i,t}, \boldsymbol{\lambda}_t\big) =: \mathcal{F}\big(\mathbf{Z}_{i,t}, \boldsymbol{\lambda}_t, R_{f,t}\big). \tag{5}$$

In this equation, the continuous function $\mathcal{F}\big(\mathbf{Z}_{i,t}, \boldsymbol{\lambda}_t, R_{f,t}\big)$ maps firm characteristics, factor risk premia, and the risk-free rate into expected returns. We further use the KA theorem (Kolmogorov 1957; Arnold 1957) to express this mapping as a flexible two-layer factor representation, providing the theoretical foundation for our empirical approach.

**Theorem (Kolmogorov 1957; Arnold 1957).** *For any integer $n \geq 2$ and any continuous function $f: [0,1]^n \to \mathbb{R}$, there exist continuous univariate functions $\Phi_q(\cdot)$ and $\phi_{q,p}(\cdot)$ such that*

$$f(x_1, \dots, x_n) = \sum_{q=1}^{2n+1} \Phi_q\left(\sum_{p=1}^{n} \phi_{q,p}\left(x_p\right)\right). \tag{6}$$

To apply the above theorem to the asset-pricing function in equation (5), we introduce the following notation. For univariate functions $\Phi_q(.)$ and $\phi_{q,p}(.)$, let $\boldsymbol{\phi}(\mathbf{Z}_{i,t}) \equiv [\phi_{q,p}(Z_{i,p,t})]_{Q\times L}$ denote the matrix-valued transformation whose $(q,p)$ element is the univariate function $\phi_{q,p}(\cdot)$ applied exclusively to characteristic $p$. For a vector $\mathbf{s} = (s_1, \dots, s_Q)'$, let $\boldsymbol{\Phi}(\mathbf{s}) = \left(\Phi_1(s_1), \dots, \Phi_Q(s_Q)\right)'$ apply the outer functions element by element. Let $\mathbf{1}_L$ and $\mathbf{1}_Q$ denote $L \times 1$ and $Q \times 1$ vectors of ones. The KA theorem implies a factor-based asset pricing model as stated in the following proposition.

**Proposition 1 (Kolmogorov–Arnold Asset Pricing Model).** *Suppose that $\mathcal{F}(\cdot)$ in equation (5) is continuous on a compact rectangle $D \subset \mathbb{R}^{L+K+1}$, and let $Q = 2(L+K+1)+1$. Then there exist continuous univariate functions $\Phi_q(\cdot)$, $\phi_{q,p}(\cdot)$, $\psi_{q,k}(\cdot)$, and $\chi_q(\cdot)$, $q = 1, \dots, Q$, common across stocks and over time, such that conditional expected returns admit the two-layer representation*

$$
\begin{aligned}
E_t[r_{i,t+1}] &= \sum_{q=1}^{Q} \Phi_q\left(\sum_{p=1}^{L} \phi_{q,p}\left(Z_{i,p,t}\right) + \sum_{k=1}^{K} \psi_{q,k}\left(\lambda_{k,t}\right) + \chi_q\left(R_{f,t}\right)\right) \\
&= \mathbf{1}_Q{}' \,\boldsymbol{\Phi}\left(\underbrace{\boldsymbol{\phi}(\mathbf{Z}_{i,t})\,\mathbf{1}_L + \boldsymbol{\delta}_t}_{\mathbf{s}_{i,t}\text{: inner layer}}\right) \\
&= \mathbf{1}_Q{}' \,\underbrace{\boldsymbol{\Phi}(\mathbf{s}_{i,t})}_{\text{outer layer}},
\end{aligned}
\tag{7}
$$

*where $\boldsymbol{\delta}_t = (\delta_{1,t}, \dots, \delta_{Q,t})'$ collects the common components across stocks, $\delta_{q,t} = \sum_{k=1}^{K} \psi_{q,k}\left(\lambda_{k,t}\right) + \chi_q\left(R_{f,t}\right)$, and $\mathbf{s}_{i,t} = \boldsymbol{\phi}(\mathbf{Z}_{i,t})\,\mathbf{1}_L + \boldsymbol{\delta}_t \in \mathbb{R}^Q$ is the vector of synthetic factors, with typical element $s_{i,q,t} = \sum_{p=1}^{L} \phi_{q,p}\left(Z_{i,p,t}\right) + \delta_{q,t}$.*

*Proof.* Since $\mathcal{F}(\cdot)$ is continuous on a compact rectangle, the KA theorem applies after an affine normalization of its arguments, yielding the first line of equation (7). Because $\boldsymbol{\lambda}_t$ and $R_{f,t}$ are common across stocks, their contributions can be collected into $\delta_{q,t}$. Expressing the result in matrix form gives the second and third lines. □

Proposition 1 decomposes expected returns into two layers, which we refer to as the KAPM. In the inner layer, each characteristic enters each synthetic factor through its own potentially nonlinear transformation $\phi_{q,p}(\cdot)$, and postmultiplying by $\mathbf{1}_L$ aggregates these transformations additively across the $L$ characteristics. Adding the common component $\boldsymbol{\delta}_t$ yields the synthetic factors $\mathbf{s}_{i,t}$. In the outer layer, $\boldsymbol{\Phi}(\cdot)$ maps these factors into conditional expected returns through a second set of univariate functions. While the inner layer preserves additivity across individual characteristics, a nonlinear outer layer allows interactions among characteristics to emerge in the overall pricing function.

The factor risk premia and the risk-free rate enter expected returns only through $\boldsymbol{\delta}_t$, which is common across stocks but may vary over time. Thus, the only stock-specific inputs to expected returns are the transformed firm characteristics $\boldsymbol{\phi}(\mathbf{Z}_{i,t})$. Under a linear outer layer, $\boldsymbol{\delta}_t$ changes the expected returns of all stocks by the same amount. Under a nonlinear outer layer, however, it can also affect cross-sectional differences in expected returns. The two-layer structure also provides a natural basis for distinguishing the contributions of nonlinear characteristic transformations and nonlinear pricing functions, which we examine through layer-wise ablation analysis.

## B. Estimation with Kolmogorov–Arnold Networks

We estimate KAPM using the Kolmogorov–Arnold Network (KAN) of Liu et al. (2025). Conventional neural networks (NNs) and KAN are motivated by two different mathematical theorems: the Universal Approximation Theorem (Cybenko 1989; Hornik, Stinchcombe, and White 1989) and the Kolmogorov–Arnold Representation Theorem (Kolmogorov 1957; Arnold 1957), respectively. Both provide mathematical foundations for representing nonlinear multivariate functions, but through different functional structures. KAN provides a machine-learning implementation of the KA theorem, making it particularly suitable for estimating the two-layer representation of expected returns established in Proposition 1.[6]

[6] In the terminology of Liu et al. (2025), NNs place fixed activation functions on nodes and learnable linear weights on edges, whereas KANs place learnable univariate functions on edges and use summation at nodes. A conventional feedforward NN with one hidden layer approximates conditional expected returns as

$$\hat{E}_t[r_{i,t+1}] = \sum_{j=1}^{H} \beta_j \, \sigma\left( a_j + \sum_{p=1}^{L} w_{j,p} \, Z_{i,p,t} \right),$$

where $\sigma(\cdot)$ is a fixed activation function and the remaining coefficients are estimated from the data. This expression has the same two-layer compositional form as KAPM, but with linear inner functions and outer functions restricted to scaled and shifted versions of a common activation function. KAN instead estimates potentially nonlinear univariate functions in both layers.

To implement the KA representation, Liu et al. (2025) parameterize each learnable univariate function in a KAN, denoted generically by $g(x)$, as a linear combination of B-spline basis functions,

$$g(x) = \sum_{m=1}^{M} \theta_m B_m^{(d)}(x), \tag{8}$$

where $B_m^{(d)}(\cdot)$ denotes the $m$-th B-spline basis function of degree $d$, and $\theta_m$ is its learnable coefficient. The spline basis is constructed over a grid of input values. A finer grid allows more localized nonlinear transformations but increases the number of parameters and the risk of overfitting. Applying this approach to KAPM, we estimate both the inner-layer characteristic transformations $\phi_{q,p}(\cdot)$ and the outer-layer pricing functions $\Phi_q(\cdot)$ using learnable spline functions, with coefficients $\theta_{q,p,m}^{\phi}$ and $\theta_{q,m}^{\Phi}$ estimated separately for every function. We estimate these coefficients jointly by minimizing squared prediction errors with an L1 penalty. Further details of the B-spline construction and KAN implementation are offered in Internet Appendix Section 1.

Two aspects of our empirical implementation warrant further discussion. First, since our empirical estimation focuses on the nonlinear mapping from firm characteristics to expected returns through synthetic factors, we do not directly estimate the factor risk premia or the risk-free rate. In the one-period representation of Proposition 1, these quantities enter through $\boldsymbol{\delta}_t$, which is common across stocks. Because its constant component can be absorbed into the inner-layer spline functions, the corresponding contribution cannot be separately identified without additional restrictions. This, however, does not affect the estimated mapping: absorbing the constant component into the spline functions leaves the model's expected returns unchanged.

Second, the value of Q provided by Proposition 1 is a theoretical upper bound on the number of synthetic factors required. In our empirical estimation, we instead treat Q as a hyperparameter to be determined from the data. Section III.C describes how we determine Q and the other hyperparameters, and Internet Appendix Section 2 provides further details.

# III. Data and Empirical Design

## A. Sample and Characteristics

Our empirical analysis covers the CRSP–Compustat universe of U.S. common stocks from 1957 to 2023. Stock returns and anomaly characteristics are obtained from the Global Factor Dataset of Jensen, Kelly, and Pedersen (2023, hereafter JKP). To mitigate the well-known small-stock bias in machine-learning asset-pricing tests (e.g., Avramov, Cheng, and Metzker 2023), we exclude microcap stocks from our main sample, using the NYSE 20th-percentile size breakpoint. Specifically, we remove stocks classified as "micro" or "nano" in the JKP dataset. The original JKP library contains 153 anomaly characteristics. To ensure sufficient data availability, we require each characteristic to be observed for at least 30% of stock-month observations. This requirement eliminates one characteristic, R&D capital-to-book assets, which has substantial missing observations due to its reliance on historical R&D expenditures. Our final sample therefore contains 152 anomaly characteristics.

Following Gu, Kelly, and Xiu (2020), we rank-normalize each characteristic cross-sectionally each month to the interval $[-1,1]$. We also follow them in replacing missing characteristics with the cross-sectional median. We then interact these characteristics with four macroeconomic state variables from Goyal, Welch, and Zafirov (2024): the term spread, aggregate stock-market volatility, the Treasury-bill rate, and the default-yield spread. We use these four variables as representative measures of aggregate conditions, capturing the level and slope of interest rates, credit risk, and stock-market uncertainty.

Let $\mathbf{c}_{i,t}$ denote the vector of 152 characteristics for stock $i$ at time $t$, and let $\mathbf{x}_t$ contain the four macroeconomic state variables and a constant. We construct the input vector as $\mathbf{Z}_{i,t} = \mathbf{x}_t \otimes \mathbf{c}_{i,t}$, where $\otimes$ denotes the Kronecker product. This construction includes the 152 original characteristics and their interactions with the four macroeconomic state variables, yielding $L = 760$ inputs. The resulting inputs are bounded over our sample, consistent with the compact-domain setting considered in Proposition 1.

## B. Estimation and Out-of-Sample Design

Similar to Gu, Kelly, and Xiu (2020), we adopt an expanding-window estimation procedure to evaluate KAPM out of sample. We initially estimate the model using data from January 1957 through December 1986. Beginning in January 1987, the estimated model generates one-month-ahead stock-return forecasts. The model is re-estimated annually using all observations available through the end of the preceding year, producing out-of-sample forecasts from 1987 through 2023. Within each test year, model parameters remain fixed, while firm characteristics and macroeconomic variables are updated monthly. To improve computational efficiency, we use the anchoring procedure described in Internet Appendix Section 1.

To evaluate the economic value of these forecasts, we sort stocks into deciles each month based on their KAPM-predicted returns. We construct both value-weighted and equally weighted portfolios, rebalanced monthly. The resulting long–short portfolios buy stocks in the highest predicted-return decile and sell stocks in the lowest.

### C. Hyperparameter and Architecture Selection

To determine Q and other hyperparameters, we combine prediction-based validation with architecture-selection tests. Similar to Gu, Kelly, and Xiu (2020), we divide each estimation sample into training and validation subsets (80% and 20%) and use validation prediction errors to select the spline grid size and regularization strength. To select the architecture, we use only the initial training sample (1957–1986) to compare specifications with different numbers of synthetic factors, spline degrees, and network depths. Both the in-sample Sharpe-ratio comparison of high-minus-low portfolios and the Bayesian factor-inclusion analysis favor an architecture with eight synthetic factors, one hidden layer, and cubic splines. We refer to this selected specification as the eight-factor KAPM, or KAPM8. Internet Appendix Sections 1 and 2 provide more details on the validation procedure and architecture selection.

## IV. Out-of-Sample Performance of KAPM8

We now examine the out-of-sample performance of KAPM8 over 1987–2023 using both portfolio analysis and cross-sectional asset-pricing tests.

### A. Portfolio Performance

Table 1 compares portfolios formed on KAPM8 return forecasts with those formed on the 152 anomaly characteristics. Each month, we sort stocks into deciles and construct high-minus-low portfolios. For the anomaly benchmark, we average the 152 individual anomaly long–short portfolios. We report both value-weighted and equally weighted returns.

The KAPM8 portfolios substantially outperform the anomaly benchmark. The value-weighted KAPM8 high-minus-low portfolio earns an annualized return of 20.19%, with a Sharpe ratio of 1.38, compared with 2.39% and 0.56 for the anomaly benchmark. The corresponding equally weighted portfolio earns 22.13%, with a Sharpe ratio of 1.68, compared with 3.03% and 0.83 for the anomalies. Returns generally increase across the predicted-return deciles, indicating that KAPM8 distinguishes stocks with different subsequent returns throughout much of the forecast distribution.

Figure 1 plots the cumulative returns of the average-anomaly portfolio and the KAPM8 portfolio, together with the KAPM3 portfolio formed from the three leading synthetic factors, which we examine in Section V. Both KAPM portfolios accumulate substantially larger returns than the anomaly benchmark.

The KAPM8 portfolios also generate economically large and statistically significant risk-adjusted returns. Their monthly FF6 alphas are 1.74% and 1.78% for value-weighted and equally weighted portfolios, respectively. These results indicate that the portfolio returns are not fully explained by conventional factor exposures.

Internet Appendix Section 4.1 benchmarks KAPM8 against the neural-network portfolios of Gu, Kelly, and Xiu (2020) over the overlapping 1987–2016 period. The value-weighted KAPM8 portfolio attains a higher Sharpe ratio (1.50 versus 1.35) and information ratio (1.34 versus 1.18), although its FF6 alpha is slightly lower (1.68% versus 1.76% per month). Differences in the predictor sets, stock universes, and portfolio construction preclude a fully matched comparison.

### B. Cross-Sectional Asset Pricing Tests

The portfolio evidence shows that KAPM8 forecasts contain substantial return-predictive information. We next examine whether these forecasts also help explain the broader cross-section of average stock returns, beyond their ability to distinguish stocks at the extremes of the predicted-return distribution. Following the tradition of cross-sectional asset-pricing tests using time-series average returns (e.g., Black, Jensen, and Scholes 1972; Adrian, Etula, and Muir 2014; He, Kelly, and Manela 2017), we regress average realized returns on average KAPM8-implied expected returns at the individual-stock level, which avoids the loss of information from aggregating stocks into portfolios (Ang, Liu, and Schwarz 2020). Unlike conventional factor-pricing tests that use estimated factor exposures, our regression directly evaluates the cross-sectional relation between predicted and realized average returns.[7]

For each stock $i$, let $\bar{r}_i$ and $\bar{\hat{r}}_i^{KAPM8}$ denote the time-series averages of its realized excess returns and KAPM8-implied expected returns, respectively, computed over the months in which the stock is observed. To examine whether KAPM8 provides information beyond conventional asset-pricing models, we estimate

$$\bar{r}_i = \alpha + b\,\bar{\hat{r}}_i^{KAPM8} + \boldsymbol{\gamma}'\widehat{\boldsymbol{\beta}}_i + \varepsilon_i, \tag{9}$$

where $\widehat{\boldsymbol{\beta}}_i$ denotes stock $i$'s estimated exposures to conventional risk factors. The baseline specification excludes these exposures, while subsequent specifications control for CAPM, FF3 (Fama and French 1993), FF5 (Fama and French 2015), and FF6 factors. We then compute Fama–MacBeth (1973) $t$-statistics, following Adrian, Etula, and Muir (2014) and Borri, Chetverikov, Liu, and Tsyvinski (2024).[8] If KAPM8 correctly describes expected returns, the intercept $\alpha$ should be zero and the coefficient $b$ should be one. Since cross-sectional tests should evaluate theoretical restrictions on intercepts and slopes (Lewellen, Nagel, and Shanken 2010), we examine both coefficients.

Table 2 reports the results. The coefficient on the KAPM8-implied expected return is positive and highly significant across all specifications. The baseline estimate is 1.136, with a $t$-statistic of 4.47. The coefficient remains close to one when conventional factor exposures are included, suggesting that KAPM8 forecasts closely align with the cross-sectional variation in average realized returns.

KAPM8 also provides substantial incremental explanatory power. The FF6-only specification produces an adjusted $R^2$ of 0.5%. Adding KAPM8-implied expected returns raises the adjusted $R^2$ to 8.3%. The information in KAPM8 forecasts therefore cannot be subsumed by conventional factor exposures.

[7] A related approach evaluates characteristic-based return forecasts using month-by-month Fama–MacBeth regressions of realized returns on the forecasts (Lewellen 2015). Our test instead relates time-series average realized returns to average predicted returns across individual stocks.

[8] More specifically, we also estimate the Fama–MacBeth version of equation (9) for each month. This yields a time series of monthly estimates for the intercept and each coefficient. The Fama–MacBeth standard error of a coefficient is then calculated as the time-series standard deviation of its monthly estimates divided by the square root of the number of months. An alternative adjustment with Newey–West lags yields the same conclusions, as shown in Internet Appendix Table IA5.

However, the pricing intercept remains statistically significant even after controlling for both FF6 factors and KAPM8-implied expected returns. In the FF6-adjusted specification, the intercept equals $-0.273\%$ per month ($t = -2.32$). Thus, neither FF6 nor KAPM8 fully explains the cross-section of average stock returns.

# V. The Dimensionality of KAPM: From Eight to Three Factors

Although the preceding analysis shows that KAPM8 has significant explanatory power for the cross-section of stock returns, its eight-factor architecture is selected using only the initial training sample. A natural question is therefore whether all eight factors contain independent pricing information out of sample. To address this question, we apply Bayesian factor selection to evaluate the incremental pricing information provided by each synthetic factor. The analysis favors a more parsimonious model based on the three leading synthetic factors. We then examine the portfolio performance and cross-sectional pricing power of the resulting three-factor model.

## A. Contributions and Selection of Synthetic Factors

To begin with, we provide preliminary evidence that the eight synthetic factors do not contribute equally to the out-of-sample performance of KAPM8. Because running KAN estimations does not imply a unique ordering of the latent factors, we rank them by an importance score based on the inner- and outer-layer spline coefficients.[9]

Figure 2 plots the out-of-sample Sharpe ratios of equally weighted long–short portfolios formed using increasing numbers of leading synthetic factors. The Sharpe ratio rises sharply over the first two factors, reaches its maximum of 1.78 with three factors, and declines to 1.68 when all eight factors are included. In other words, adding further factors after the first three does not seem to improve performance, suggesting that the number of synthetic factors providing distinct return-predictive information may be substantially smaller than eight. To assess this possibility more formally, we conduct Bayesian factor selection.

[9] For factor $q$, we define its inner-layer importance as $I_q^{\text{in}} = \sum_{p=1}^{L} \sum_m \left|\theta_{q,p,m}^{\phi}\right|$ and its outer-layer importance as $I_q^{\text{out}} = \sum_m \left|\theta_{q,m}^{\Phi}\right|$, where $\theta_{q,p,m}^{\phi}$ and $\theta_{q,m}^{\Phi}$ denote the corresponding spline coefficients. We then define the overall importance score as the product of the relative inner-layer and outer-layer importance measures, $S_q = \frac{I_q^{\text{in}}}{\sum_{j=1}^{Q} I_j^{\text{in}}} \frac{I_q^{\text{out}}}{\sum_{j=1}^{Q} I_j^{\text{out}}}$, and we rank the synthetic factors in descending order of $S_q$. The ranking is performed separately in each estimation window, and the return series of each ranked factor links the factors of the same rank across annual re-estimations.

Following Barillas and Shanken (2018) and Chib, Zeng, and Zhao (2020), we evaluate all $2^8 - 1 = 255$ nonempty subsets of the eight synthetic factors, represented by their corresponding long–short portfolio returns. Assigning equal prior probabilities to all candidate models, we compute the posterior probability of each model and the posterior inclusion probability of each factor. The Bayesian analysis evaluates which factors provide information not spanned by the others while accounting for model complexity. Internet Appendix Section 3 describes the procedure in detail.

Table 3 reports the full-sample results. Panel A shows that the first two synthetic factors have posterior inclusion probabilities of 100%, while the third has an inclusion probability of 62.94%. In contrast, the remaining five factors have inclusion probabilities ranging from 11.92% to 30.36%. Panel B reports the five models with the highest posterior probabilities. The model containing the first three factors has the highest posterior probability (21.71%), followed by the model with the first two factors (12.63%). Each of the next three models adds one additional factor to the first three. These results suggest that posterior support is concentrated on the leading synthetic factors, with particularly strong support for the first two.

We further examine the stability of these results through recursive Bayesian estimation. Holding the initial 44-month prior-training sample fixed, we update posterior model and factor inclusion probabilities as additional observations become available. Figure 3 shows that the first two factors consistently exhibit inclusion probabilities close to 100%, whereas support for the third factor varies over time. Nonetheless, the full-sample Bayesian analysis favors a parsimonious specification containing the first three factors, as indicated in Table 3. We refer to this specification as KAPM3.

## B. Performance of KAPM3

We next evaluate the portfolio performance and cross-sectional pricing power of KAPM3. While the underlying factor predictions are generated out of sample, the three-factor specification is selected using posterior evidence from the full evaluation period.

Table 4 reports the performance of portfolios formed on the three leading synthetic factors jointly and individually. KAPM3 achieves annualized Sharpe ratios of 1.43 and 1.78 for value-weighted and equally weighted portfolios, respectively, compared with 1.38 and 1.68 for KAPM8. Its monthly FF6 alphas are 1.75% and 1.82%, compared with 1.74% and 1.78% for KAPM8. The combined three-factor portfolio also outperforms portfolios formed on each factor individually. Figure 1 further shows that KAPM3 accumulates higher returns than KAPM8 by the end of the evaluation period. These results suggest that much of the economic value of KAPM8 is concentrated in the leading factors.

Table 5 compares the risk and trading characteristics of these portfolios. Relative to KAPM8, KAPM3 has smaller maximum drawdowns (−28.06% versus −30.35% for value-weighted portfolios and −19.16% versus −27.07% for equally weighted portfolios). It also has slightly smaller worst monthly losses and slightly lower monthly turnover (65.3% versus 69.9% value-weighted, and 57.5% versus 60.7% equally weighted). Its returns are more positively skewed, with value-weighted skewness of 0.33 versus 0.29, but also more fat-tailed, with kurtosis of 2.48 versus 1.49. Both KAPM portfolios have substantially larger drawdowns and more than twice the turnover of the anomaly benchmark, whose turnover is about 26%. The individual factors highlight the complementarity of the leading factors: the third factor alone earns an FF6 alpha close to zero and experiences drawdowns exceeding 55%, yet the portfolio combining all three attains the highest FF6 alpha and information ratio among the KAPM portfolios. Reducing the model to its leading factors therefore lowers downside risk and modestly reduces trading intensity, while maintaining strong risk-adjusted performance.

We next examine the cross-sectional pricing power of these factors. Recall from equation (7) that the KAPM forecast can be decomposed into individual synthetic-factor contributions, $\hat{E}_t\left[r_{i,t+1}\right] = \sum_{q=1}^{Q} \Phi_q\left(s_{i,q,t}\right)$. For each stock $i$, let $\overline{\Phi}_{q,i}$denote the time-series average of $\Phi_q\left(s_{i,q,t}\right)$, the contribution of synthetic factor $q$ to its predicted expected return. Replacing the aggregate KAPM8 prediction in equation (9) with the three leading factor contributions, we estimate

$$\bar{r}_i = \alpha + \sum_{q=1}^{3} b_q\, \overline{\Phi}_{q,i} + \boldsymbol{\gamma}'\widehat{\boldsymbol{\beta}}_i + \varepsilon_i. \tag{10}$$

Here, $\widehat{\boldsymbol{\beta}}_i$ denotes stock $i$'s exposures to conventional risk factors, as in Section IV.B.

Table 6 reports the results. All three synthetic factors exhibit significant cross-sectional explanatory power across all specifications, with the first factor highly significant and the other two marginally significant at the 10% level. Because the three factor contributions enter jointly, the coefficient on each factor measures its partial cross-sectional association with average returns, conditional on the contributions of the other two. These results persist after controlling for CAPM, FF3, FF5, and FF6 factor exposures. When the three factors enter jointly, the cross-sectional intercept becomes economically small and statistically insignificant. In the FF6-adjusted specification, the intercept is 0.141% per month ($t = 0.22$), compared with 0.544% ($t = 2.86$) for FF6 alone and $-0.273\%$ ($t = -2.32$) for KAPM8 in Table 2. Column (6) further replaces the three contributions with the aggregate KAPM3-implied expected return, mirroring the KAPM8 specification in Table 2. The coefficient is 1.150 (t=2.93), and the intercept is economically small and statistically insignificant (0.048%, t=0.19).

We further examine whether KAPM3 contains information beyond existing asset-pricing factors. Table 7 compares KAPM3 with established economic factors, including the q-factor model, traded liquidity, quality-minus-junk, betting-against-beta, and mispricing factors, as well as statistical factor models based on PCA, RP-PCA, kernel PCA, and IPCA. For each benchmark, we first estimate cross-sectional regressions using the benchmark factor exposures and then add the three KAPM3 contributions. The benchmark specifications produce significant cross-sectional intercepts when estimated without KAPM3. Adding KAPM3 renders these intercepts statistically insignificant across all specifications, while the three synthetic factors retain explanatory power at least at the 10% level. Internet Appendix Section 4.2 reports additional cross-sectional tests using these benchmarks.

Finally, we compare KAPM3 with the 152 individual anomaly factors from which the synthetic factors are constructed. For each anomaly factor $k$, denoted by $A_{k,t}$, we first estimate stock-level exposures from the time-series regression

$$r_{i,t+1} = a_{i,k} + \beta^{A}_{i,k} A_{k,t+1} + u_{i,k,t+1}. \tag{11}$$

We then estimate a cross-sectional regression using each stock's average realized excess return:

$$\bar{r}_i = \alpha_k + d_k \, \hat{\beta}^{A}_{i,k} + \varepsilon_{i,k}. \tag{12}$$

To examine the incremental explanatory power of KAPM3, we augment this regression with the three leading synthetic-factor contributions:

$$\bar{r}_i = \alpha_k + d_k \, \hat{\beta}^{A}_{i,k} + \sum_{q=1}^{3} b_{q,k} \, \overline{\Phi}_{q,i} + \varepsilon_{i,k}. \tag{13}$$

Comparing equations (12) and (13) allows us to assess whether KAPM3 explains average returns beyond the information contained in each individual anomaly factor.

Figure 4 summarizes the distribution of absolute t-statistics across the 152 anomaly specifications. When anomaly exposures enter alone, about three-quarters of the cross-sectional intercepts are statistically significant at the 10% level, with a mean absolute t-statistic of 1.89 and a median of 1.75. In contrast, the anomaly coefficients are rarely significant: their absolute t-statistics have a mean of only 0.61 and reach just 1.17 even at the 90th percentile.

After including the synthetic factors of KAPM3, the intercepts become insignificant, with a mean absolute t-statistic of 0.34 and only 0.56 at the 90th percentile. The three synthetic factors, by comparison, are significant: even at the 10th percentile, their absolute t-statistics are 2.80, 2.09, and 1.93, respectively. The anomaly coefficients remain insignificant, with a 90th-percentile absolute t-statistic of 1.39. These results suggest that KAPM3 captures cross-sectional return information beyond the individual anomalies used to construct it.

Taken together, the results indicate that three synthetic factors provide a parsimonious representation of the return-predictive information embedded in the factor zoo. The leading factors account for much of KAPM8's portfolio performance and exhibit cross-sectional explanatory power beyond existing factor models and individual anomalies.

## VI. Sources of Nonlinearity and Economic Interpretation

The preceding sections show that KAPM explains the cross-section of stock returns with a small number of synthetic factors. Beyond this explanatory power, the two-layer structure of KAPM has two important asset-pricing implications for understanding the factor zoo. The first concerns the sources of nonlinearity. Because the inner layer allows nonlinear characteristic aggregation and the outer layer allows nonlinear factor pricing, restricting each layer to be linear in turn reveals which source of nonlinearity accounts for KAPM's portfolio performance and cross-sectional pricing power. The second concerns the economic content of the synthetic factors. A low-dimensional factor structure may arise because only a few anomalies contain pricing-relevant information, or because information dispersed across many anomalies is aggregated into a few factors. In this section, we examine the first implication through layer-wise ablation and the second by tracing the leading synthetic factors back to the underlying anomalies and their economic themes.

## A. Sources of Nonlinearity: Layer-Wise Ablation

The KAPM architecture separates two potential sources of nonlinearity. The inner-layer functions, $\phi_{q,p}(\cdot)$, transform firm characteristics into synthetic factors, while the outer-layer functions, $\Phi_q(\cdot)$, map these factors into predicted returns. We examine whether KAPM's predictive gains arise primarily from nonlinear characteristic aggregation, nonlinear factor pricing, or their combination.

These two channels correspond to distinct approaches in the asset-pricing literature. Nonlinear transformations of individual characteristics are central to the nonparametric analysis of Freyberger, Neuhierl, and Weber (2020). With a linear outer layer, KAPM reduces to an additive function of characteristics, $\sum_p g_p\left(Z_{i,p,t}\right)$, and thus nests their additive nonparametric model. Conversely, a linear inner layer combined with nonlinear outer-layer functions allows nonlinear pricing of linear characteristic indices, in the spirit of Bansal and Viswanathan (1993) and Rosenberg and Engle (2002). KAPM allows both forms of nonlinearity to operate jointly.

To assess their respective contributions, we consider four combinations of linear and nonlinear transformations in the two layers. The L/L specification restricts both layers to affine functions, yielding a linear characteristic–return mapping. The NL/L specification allows nonlinear characteristic aggregation but imposes linear pricing of the resulting synthetic factors. The L/NL specification constructs synthetic factors as linear combinations of characteristics but allows nonlinear pricing functions. The NL/NL specification retains the full flexibility of KAPM.

We implement these restrictions by modifying the estimated KAPM models without retraining them. Specifically, we replace selected spline functions with their global linear approximations, fitted by least squares over a dense grid of their empirical support, while holding the remaining functions fixed. This layer-wise ablation allows us to assess the contribution of each source of nonlinearity within the same estimated model. We also include a linear PCA benchmark based on the first three principal components of anomaly portfolio returns. Table 8 reports the portfolio performance of the resulting specifications for both KAPM8 and KAPM3.

The linear PCA benchmark performs substantially below the nonlinear KAPM specifications, with annualized Sharpe ratios of 0.52 and 0.69 for value-weighted and equally weighted portfolios, respectively, compared with 1.38 and 1.68 for the unrestricted KAPM8. The restricted specifications also retain substantial portfolio performance. For KAPM8, NL/L achieves annualized Sharpe ratios of 1.19 and 1.47 for value-weighted and equally weighted portfolios, respectively, while L/NL achieves 1.25 and 1.55. The corresponding Sharpe ratios for NL/L KAPM3 are 1.26 and 1.54.

Relative to the PCA benchmark, NL/L captures 78% of the value-weighted and 79% of the equally weighted Sharpe-ratio improvement achieved by the unrestricted KAPM8, and 81% and 78% of that achieved by the unrestricted KAPM3. The NL/L specifications also earn monthly FF6 alphas of 1.36% and 1.52% for KAPM8 and 1.45% and 1.58% for KAPM3, compared with 0.76% and 0.89% for the PCA benchmark. Thus, both sources of nonlinearity contribute to portfolio performance, and much of KAPM's performance survives when either layer is restricted to be linear.

The NL/L specification also provides a comparison with Freyberger, Neuhierl, and Weber (2020). Over 1991–2014, their equally weighted long–short portfolios formed on nonparametric and linear return forecasts achieve out-of-sample Sharpe ratios of 2.75 and 1.06, respectively. Excluding stocks below the NYSE 20th size percentile reduces these Sharpe ratios to 0.89 and 0.06. In our size-screened sample, the NL/L specifications achieve equally weighted Sharpe ratios of 1.47 for KAPM8 and 1.54 for KAPM3 over 1987–2023. Differences in sample periods, predictors, and estimation procedures prevent a direct performance comparison, but the results highlight the predictive value of nonlinear characteristic transformations.

Relatedly, Lu, Spiegel, and Zhang (2025) develop a dynamic arbitrage strategy in which expected arbitrage profits are additive in squared anomaly alphas. Using 152 firm characteristics, they show that this economically motivated strategy accounts for approximately 87.9 basis points of the monthly alphas of neural-network high-minus-low portfolios, and fully explains neural-network performance once unpublished anomalies and microcap stocks are excluded. Their findings provide further evidence that additive nonlinear effects can account for substantial return predictability, consistent with the strong performance of our NL/L specification.

Taken together, the portfolio results suggest that both sources of nonlinearity contribute to KAPM's predictive performance. However, the strong performance of NL/L indicates that nonlinear characteristic transformations alone can preserve much of the predictive information captured by the unrestricted model. This raises a further question: whether these transformations are also sufficient to explain the cross-section of average stock returns.

To address this question, we examine the cross-sectional pricing power of the NL/L specifications and report the results in Table 9. Using the aggregate return prediction from KAPM3, the coefficient is 0.837 (t=4.33), and the FF6-adjusted cross-sectional intercept is only 0.030% per month (t=0.14). When the three synthetic-factor contributions enter separately, the intercept remains insignificant at −0.136% (t=−0.57), while all three factors exhibit significant explanatory power at least at the 10% level. In contrast, the NL/L KAPM8 specification leaves a significant intercept of 0.362% (t=2.49). These results suggest that nonlinear characteristic aggregation, even without nonlinear pricing of the synthetic factors, is sufficient for KAPM3 to eliminate the pricing intercept that traditional models leave unexplained.

The recovered outer-layer functions help explain why linear pricing of the synthetic factors preserves most of the performance of KAPM. As shown in Internet Appendix Figure IA2, the pricing function of the first factor is approximately linear, and that of the second is monotonic with moderate curvature in the tails. The pricing function of the third factor is U-shaped, but its variation is much smaller in magnitude than that of the first two.

Overall, our results suggest that additive nonlinear transformations of individual characteristics are sufficient to achieve most of the portfolio performance of KAPM and help explain the cross-section of average stock returns in our tests, without requiring interactions among characteristics or nonlinear pricing of the synthetic factors.

### B. Economic Content of the Synthetic Factors

Two approaches in the asset-pricing literature seek to explain the common structure underlying the factor zoo. The first identifies a small set of economically motivated factors that can explain a broad range of anomaly returns (e.g., Fama and French 2015; Hou, Xue, and Zhang 2015; Hou, Mo, Xue, and Zhang 2021). The second extracts pricing-relevant information from a large universe of anomaly factors through statistical dimension reduction and regularization (e.g., Kozak, Nagel, and Santosh 2018, 2020). The latter approach allows pricing information to be distributed across many anomalies rather than concentrated in a small subset of existing factors. KAPM allows us to distinguish between these two perspectives, because each synthetic factor can be traced back to the underlying anomalies. We therefore examine the characteristics from which the three leading synthetic factors are constructed.

We use the inner-layer spline-coefficient importance measure underlying Section V.A to assess the contribution of each of the 760 inputs to each synthetic factor. We then aggregate the scores of each anomaly and its macroeconomic interactions and group the 152 anomalies into the 13 economic themes classified by Jensen, Kelly, and Pedersen (2023). Within each theme, we average across anomalies, so that themes containing more anomalies do not receive greater weight mechanically. Figure 5 reports the resulting theme composition of the three leading factors.

The three factors draw on overlapping but distinct sets of anomaly themes. Momentum receives the greatest weight in the first and second factors, while low-risk characteristics are most important for the third. Short-term reversal also contributes prominently to the first and third factors, whereas low-risk is particularly important for the second. More importantly, no single theme dominates any of the three factors: the largest theme share is about 13% for the first factor and about 10% for the second and the third, compared with 7.7% if the thirteen themes were weighted equally. These findings support the second perspective: KAPM achieves a low-dimensional representation by synthesizing information across a broad set of anomalies, rather than by selecting a small subset of existing factors.

# VII. Conclusion

This paper examines the dimensionality of return-predictive information embedded in firm characteristics and the sources of nonlinearity in the mapping from characteristics to expected returns. We develop the Kolmogorov–Arnold Asset Pricing Model (KAPM), a two-layer asset-pricing framework based on the Kolmogorov–Arnold representation theorem, and estimate it using Kolmogorov–Arnold Networks. The inner layer aggregates firm characteristics into synthetic factors, while the outer layer maps these factors into expected returns. This structure allows us to examine both the number of synthetic factors needed to summarize return-predictive information and the nonlinear transformations through which that information enters expected returns.

Applying the framework to the factor zoo, we find that a model with eight synthetic factors delivers strong out-of-sample portfolio performance and cross-sectional explanatory power. Bayesian selection over the full evaluation period favors a more parsimonious three-factor specification, while recursive estimates provide more consistent support for the leading two factors. The three-factor specification delivers Sharpe ratios and FF6 alphas at least as high as those of the full model. Its three factor contributions jointly produce an insignificant cross-sectional intercept and provide explanatory power beyond conventional asset-pricing factors and individual anomalies.

Beyond its pricing performance, KAPM provides two important asset-pricing implications for understanding the factor zoo. First, layer-wise ablation shows that nonlinear characteristic aggregation and nonlinear factor pricing both contribute to portfolio performance. However, additive nonlinear transformations of individual characteristics suffice to capture much of KAPM's portfolio performance and explain the cross-section of average stock returns in the KAPM3 specification, without interactions across characteristics. Second, the leading synthetic factors combine information from multiple anomaly themes rather than selecting a small subset of existing factors. More broadly, our findings suggest that the hidden structure of the factor zoo is both low-dimensional and nonlinear, with a small number of synthetic factors capturing information distributed across a broad set of characteristics.

## References

Adrian, Tobias, Erkko Etula, and Tyler Muir, 2014, Financial intermediaries and the cross-section of asset returns, Journal of Finance 69, 2557–2596.

Aït-Sahalia, Yacine, and Andrew W. Lo, 2000, Nonparametric risk management and implied risk aversion, Journal of Econometrics 94, 9–51.

Ang, Andrew, Jun Liu, and Krista Schwarz, 2020, Using stocks or portfolios in tests of factor models, Journal of Financial and Quantitative Analysis 55, 709–750.

Arnold, Vladimir I., 1957, On functions of three variables, Doklady Akademii Nauk SSSR 114, 679–681.

Asness, Clifford S., Andrea Frazzini, and Lasse Heje Pedersen, 2019, Quality minus junk, Review of Accounting Studies 24, 34–112.

Avramov, Doron, Si Cheng, and Lior Metzker, 2023, Machine learning vs. economic restrictions: Evidence from stock return predictability, Management Science 69, 2587–2619.

Bansal, Ravi, and S. Viswanathan, 1993, No arbitrage and arbitrage pricing: A new approach, Journal of Finance 48, 1231–1262.

Barillas, Francisco, and Jay Shanken, 2018, Comparing asset pricing models, Journal of Finance 73, 715–754.

Bell, Sebastian, Ali Kakhbod, Martin Lettau, and Abdolreza Nazemi, 2026, Glass box machine learning and corporate bond returns, Journal of Financial Economics 181, 104294.

Black, Fischer, Michael C. Jensen, and Myron Scholes, 1972, The capital asset pricing model: Some empirical tests, in Michael C. Jensen, ed., Studies in the Theory of Capital Markets (Praeger, New York).

Borri, Nicola, Denis Chetverikov, Yukun Liu, and Aleh Tsyvinski, 2024, One factor to bind the cross-section of returns, NBER Working Paper No. 32365.

Bryzgalova, Svetlana, Jiantao Huang, and Christian Julliard, 2023, Bayesian solutions for the factor zoo: We just ran two quadrillion models, Journal of Finance 78, 487–557.

Bryzgalova, Svetlana, Markus Pelger, and Jason Zhu, 2025, Forest through the trees: Building cross-sections of stock returns, Journal of Finance 80, 2447–2506.

Carhart, Mark M., 1997, On persistence in mutual fund performance, Journal of Finance 52, 57–82.

Chamberlain, Gary, and Michael Rothschild, 1983, Arbitrage, factor structure, and mean–variance analysis on large asset markets, Econometrica 51, 1281–1304.

Chen, Luyang, Markus Pelger, and Jason Zhu, 2024, Deep learning in asset pricing, Management Science 70, 714–750.

Chib, Siddhartha, Ximing Zeng, and Lingxiao Zhao, 2020, On comparing asset pricing models, Journal of Finance 75, 551–577.

Cochrane, John H., 2011, Presidential address: Discount rates, Journal of Finance 66, 1047–1108.

Cong, Lin William, Guanhao Feng, Jingyu He, and Xin He, 2025, Growing the efficient frontier on panel trees, Journal of Financial Economics 167, 104024.

Connor, Gregory, Matthias Hagmann, and Oliver Linton, 2012, Efficient semiparametric estimation of the Fama–French model and extensions, Econometrica 80, 713–754.

Cybenko, George, 1989, Approximation by superpositions of a sigmoidal function, Mathematics of Control, Signals and Systems 2, 303–314.

Didisheim, Antoine, Shikun (Barry) Ke, Bryan T. Kelly, and Semyon Malamud, 2023, Complexity in factor pricing models, NBER Working Paper No. 31689.

Didisheim, Antoine, Shikun (Barry) Ke, Bryan T. Kelly, and Semyon Malamud, 2024, The surprising dominance of large factor models, NBER Working Paper No. 33012.

Fama, Eugene F., and Kenneth R. French, 1993, Common risk factors in the returns on stocks and bonds, Journal of Financial Economics 33, 3–56.

Fama, Eugene F., and Kenneth R. French, 2015, A five-factor asset pricing model, Journal of Financial Economics 116, 1–22.

Fama, Eugene F., and James D. MacBeth, 1973, Risk, return, and equilibrium: Empirical tests, Journal of Political Economy 81, 607–636.

Feng, Guanhao, Stefano Giglio, and Dacheng Xiu, 2020, Taming the factor zoo: A test of new factors, Journal of Finance 75, 1327–1370.

Frazzini, Andrea, and Lasse Heje Pedersen, 2014, Betting against beta, Journal of Financial Economics 111, 1–25.

Freyberger, Joachim, Andreas Neuhierl, and Michael Weber, 2020, Dissecting characteristics nonparametrically, Review of Financial Studies 33, 2326–2377.

Goyal, Amit, Ivo Welch, and Athanasse Zafirov, 2024, A comprehensive look at the empirical performance of equity premium prediction II, Review of Financial Studies 37, 3490–3557.

Gu, Shihao, Bryan T. Kelly, and Dacheng Xiu, 2020, Empirical asset pricing via machine learning, Review of Financial Studies 33, 2223–2273.

Gu, Shihao, Bryan T. Kelly, and Dacheng Xiu, 2021, Autoencoder asset pricing models, Journal of Econometrics 222, 429–450.

Harvey, Campbell R., Yan Liu, and Heqing Zhu, 2016, … and the cross-section of expected returns, Review of Financial Studies 29, 5–68.

He, Zhiguo, Bryan Kelly, and Asaf Manela, 2017, Intermediary asset pricing: New evidence from many asset classes, Journal of Financial Economics 126, 1–35.

Hornik, Kurt, 1991, Approximation capabilities of multilayer feedforward networks, Neural Networks 4, 251–257.

Hornik, Kurt, Maxwell Stinchcombe, and Halbert White, 1989, Multilayer feedforward networks are universal approximators, Neural Networks 2, 359–366.

Hou, Kewei, Haitao Mo, Chen Xue, and Lu Zhang, 2021, An augmented q-factor model with expected growth, Review of Finance 25, 1–41.

Hou, Kewei, Chen Xue, and Lu Zhang, 2015, Digesting anomalies: An investment approach, Review of Financial Studies 28, 650–705.

Hou, Kewei, Chen Xue, and Lu Zhang, 2020, Replicating anomalies, Review of Financial Studies 33, 2019–2133.

Jensen, Theis Ingerslev, Bryan T. Kelly, and Lasse Heje Pedersen, 2023, Is there a replication crisis in finance?, Journal of Finance 78, 2465–2518.

Kelly, Bryan T., Semyon Malamud, and Kangying Zhou, 2024, The virtue of complexity in return prediction, Journal of Finance 79, 459–503.

Kelly, Bryan T., Seth Pruitt, and Yinan Su, 2019, Characteristics are covariances: A unified model of risk and return, Journal of Financial Economics 134, 501–524.

Kelly, Bryan T., and Dacheng Xiu, 2023, Financial machine learning, Foundations and Trends in Finance 13, 205–363.

Kolmogorov, Andrey N., 1957, On the representation of continuous functions of many variables by superposition of continuous functions of one variable and addition, Doklady Akademii Nauk SSSR 114, 953–956.

Kozak, Serhiy, Stefan Nagel, and Shrihari Santosh, 2018, Interpreting factor models, Journal of Finance 73, 1183–1223.

Kozak, Serhiy, Stefan Nagel, and Shrihari Santosh, 2020, Shrinking the cross-section, Journal of Financial Economics 135, 271–292.

Lettau, Martin, and Markus Pelger, 2020, Factors that fit the time series and cross-section of stock returns, Review of Financial Studies 33, 2274–2325.

Lewellen, Jonathan, 2015, The cross-section of expected stock returns, Critical Finance Review 4, 1–44.

Lewellen, Jonathan, Stefan Nagel, and Jay Shanken, 2010, A skeptical appraisal of asset pricing tests, Journal of Financial Economics 96, 175–194.

Li, Bin, Alberto G. Rossi, Xuemin (Sterling) Yan, and Lingling Zheng, 2025, Machine learning from a “universe” of signals: The role of feature engineering, Journal of Financial Economics 172, 104138.

Liu, Ziming, Yixuan Wang, Sachin Vaidya, Fabian Ruehle, James Halverson, Marin Soljačić, Thomas Y. Hou, and Max Tegmark, 2025, KAN: Kolmogorov–Arnold networks, in The Thirteenth International Conference on Learning Representations (ICLR).

Lu, Huahao, Matthew I. Spiegel, and Hong Zhang, 2025, Machine learning as arbitrage: The economics behind neural network portfolio selection, SSRN Working Paper No. 5179857.

McLean, R. David, and Jeffrey Pontiff, 2016, Does academic research destroy stock return predictability?, Journal of Finance 71, 5–32.

Murray, Scott, Yusen Xia, and Houping Xiao, 2024, Charting by machines, Journal of Financial Economics 153, 103791.

Pástor, Ľuboš, and Robert F. Stambaugh, 2003, Liquidity risk and expected stock returns, Journal of Political Economy 111, 642–685.

Rosenberg, Joshua V., and Robert F. Engle, 2002, Empirical pricing kernels, Journal of Financial Economics 64, 341–372.

Schölkopf, Bernhard, Alexander Smola, and Klaus-Robert Müller, 1998, Nonlinear component analysis as a kernel eigenvalue problem, Neural Computation 10, 1299–1319.

Stambaugh, Robert F., and Yu Yuan, 2017, Mispricing factors, Review of Financial Studies 30, 1270–1315.

Wang, Tianqi, and Shubham Singh, 2024, KAN based autoencoders for factor models, Working paper, arXiv:2408.02694.

## Table 1: Performance of Anomaly-Based and KAPM8 Portfolios

This table reports summary statistics for anomaly-based and KAPM8 portfolios, denoted "Anomalies" and "KAPM8," respectively. For both approaches, we first compute decile-wise returns by sorting stocks into ten portfolios each month. The high-minus-low (HML) strategy is the tenth minus the first decile return, marked "10−1" in the table. For the anomaly benchmark, this HML return is computed for each of the 152 individual anomaly characteristics, and the aggregate portfolio is the simple average across them. The KAPM8 portfolio follows an identical HML construction based on KAN's return predictions. The sample covers 1987–2023 and excludes micro- and nano-cap stocks following the JKP size classification. Returns and volatilities are annualized and reported in percent. We report value-weighted (Panel A) and equal-weighted (Panel B) results.

| | Anomalies | | | | KAPM8 | | | |
|---|---|---|---|---|---|---|---|---|
| Decile | Ret | T-stat | Std | SR | Ret | T-stat | Std | SR |
| **Panel A: Value-Weighted** | | | | | | | | |
| 10 | 8.61 | 3.43 | 15.27 | 0.56 | 19.50 | 5.85 | 20.24 | 0.89 |
| 9 | 8.40 | 3.53 | 14.47 | 0.58 | 13.15 | 4.39 | 18.21 | 0.68 |
| 8 | 8.31 | 3.52 | 14.34 | 0.58 | 10.95 | 3.92 | 16.99 | 0.61 |
| 7 | 8.15 | 3.42 | 14.50 | 0.56 | 12.80 | 4.82 | 16.12 | 0.75 |
| 6 | 7.93 | 3.17 | 15.20 | 0.52 | 7.76 | 3.11 | 15.15 | 0.49 |
| 5 | 7.86 | 3.09 | 15.47 | 0.51 | 7.75 | 3.11 | 15.13 | 0.49 |
| 4 | 7.75 | 3.14 | 15.00 | 0.52 | 7.36 | 2.96 | 15.09 | 0.47 |
| 3 | 7.47 | 2.96 | 15.32 | 0.49 | 7.11 | 3.01 | 14.33 | 0.48 |
| 2 | 7.19 | 2.73 | 16.03 | 0.45 | 2.71 | 1.10 | 14.93 | 0.18 |
| 1 | 6.22 | 2.09 | 18.02 | 0.34 | -0.69 | -0.21 | 19.57 | -0.04 |
| 10−1 | 2.39 | 3.38 | 4.30 | 0.56 | 20.19 | 9.09 | 13.50 | 1.38 |
| FF6 α | 0.21 | 2.82 | - | - | 1.74 | 6.01 | - | - |
| **Panel B: Equal-Weighted** | | | | | | | | |
| 10 | 10.47 | 3.42 | 18.62 | 0.56 | 19.90 | 5.60 | 21.59 | 0.85 |
| 9 | 10.46 | 3.61 | 17.60 | 0.59 | 15.02 | 4.63 | 19.71 | 0.71 |
| 8 | 10.25 | 3.59 | 17.33 | 0.59 | 13.64 | 4.41 | 18.80 | 0.68 |
| 7 | 9.98 | 3.49 | 17.38 | 0.57 | 13.01 | 4.39 | 18.00 | 0.68 |
| 6 | 9.07 | 3.06 | 18.02 | 0.50 | 10.20 | 3.62 | 17.12 | 0.57 |
| 5 | 9.20 | 3.09 | 18.10 | 0.51 | 9.15 | 3.23 | 17.19 | 0.51 |
| 4 | 9.71 | 3.31 | 17.81 | 0.55 | 9.13 | 3.34 | 16.61 | 0.53 |
| 3 | 9.51 | 3.20 | 18.06 | 0.53 | 8.00 | 2.89 | 16.82 | 0.46 |
| 2 | 9.07 | 2.93 | 18.84 | 0.48 | 5.02 | 1.71 | 17.85 | 0.27 |
| 1 | 7.44 | 2.14 | 21.28 | 0.35 | -2.23 | -0.56 | 24.22 | -0.09 |
| 10−1 | 3.03 | 5.04 | 3.65 | 0.83 | 22.13 | 11.01 | 12.21 | 1.68 |
| FF6 α | 0.25 | 4.55 | - | - | 1.78 | 6.24 | - | - |

## Table 2: Cross-Sectional Asset-Pricing Tests: KAPM8 Factor

This table reports cross-sectional asset-pricing tests using individual stocks. For each stock, the dependent variable is the time-series average of realized excess returns, and the KAPM signal is the time-series average of the model-implied conditional expected return. Columns (1)– (4) add the KAPM signal and progressively control for stock-level exposures to the CAPM, FF3, and FF5 models. Column (5) is a benchmark that includes only the FF6 factor exposures, with no KAPM signal, so that readers can gauge how much of the cross-section traditional factors explain on their own. Column (6) adds the KAPM signal to the same FF6 exposures. Regressions use all stocks with at least twelve-monthly observations. Fama–MacBeth *t*-statistics are reported in parentheses. The standard error of a coefficient is the time-series standard deviation of its monthly estimates divided by the square root of the number of months. *, **, *** denote significance at the 10%, 5%, and 1% levels.

| | (1)<br>Bare | (2)<br>CAPM | (3)<br>FF3 | (4)<br>FF5 | (5)<br>FF6 (No KAPM) | (6)<br>FF6 |
|---|---|---|---|---|---|---|
| KAPM8 | 1.136*** | 1.145*** | 1.199*** | 1.168*** | | 1.161*** |
| | (4.47) | (4.88) | (5.61) | (5.33) | | (5.19) |
| MKT−RF | | 0.038 | 0.154 | 0.143 | 0.016 | 0.101 |
| | | (0.14) | (0.60) | (0.66) | (0.08) | (0.51) |
| SMB | | | -0.106 | -0.088 | -0.140 | -0.045 |
| | | | (-0.50) | (-0.47) | (-0.82) | (-0.27) |
| HML | | | -0.270* | -0.157 | -0.013 | -0.086 |
| | | | (-1.67) | (-1.09) | (-0.08) | (-0.57) |
| RMW | | | | -0.008 | 0.086 | 0.003 |
| | | | | (-0.07) | (0.81) | (0.03) |
| CMA | | | | -0.080 | -0.037 | -0.072 |
| | | | | (-0.85) | (-0.40) | (-0.80) |
| MOM | | | | | 0.081 | 0.112 |
| | | | | | (0.54) | (0.75) |
| α | -0.207** | -0.254** | -0.261** | -0.282** | 0.544*** | -0.273** |
| | (-2.31) | (-2.01) | (-2.36) | (-2.48) | (2.86) | (-2.32) |
| Adj. $R^2$ | 0.080 | 0.080 | 0.095 | 0.087 | 0.005 | 0.083 |
| $R^2$ | 0.080 | 0.080 | 0.096 | 0.088 | 0.007 | 0.085 |
| Months | 443 | 443 | 443 | 443 | 443 | 443 |
| Stocks | 3910 | 3910 | 3910 | 3910 | 3910 | 3910 |

## Table 3: Bayesian Selection of KAPM Synthetic Factors

This table reports Bayesian factor-selection results over the universe of eight KAPM synthetic factors. We evaluate all $2^8-1 = 255$ nonempty factor combinations and compute posterior model probabilities from the marginal likelihood of each candidate model. Panel A reports the posterior inclusion probability of each synthetic factor, defined as the sum of posterior probabilities across all models that contain the factor. Panel B reports the five models with the highest posterior probabilities, where $F_q$ denotes the $q$-th ranked synthetic factor.

Panel A: Factor Inclusion Probabilities

| Factor | Inclusion Probability |
|---|---|
| $F_1$ | 100.00% |
| $F_2$ | 100.00% |
| $F_3$ | 62.94% |
| $F_4$ | 11.92% |
| $F_5$ | 19.95% |
| $F_6$ | 13.13% |
| $F_7$ | 24.84% |
| $F_8$ | 30.36% |

Panel B: Top 5 Models by Posterior Probability

| Rank | Risk Factors | Posterior Probability |
|---|---|---|
| 1 | $F_1, F_2, F_3$ | 21.713% |
| 2 | $F_1, F_2$ | 12.627% |
| 3 | $F_1, F_2, F_3, F_8$ | 7.882% |
| 4 | $F_1, F_2, F_3, F_5$ | 5.611% |
| 5 | $F_1, F_2, F_3, F_7$ | 5.141% |

## Table 4: Performance of KAPM Portfolios Based on the Leading Factors

This table reports decile-sorted portfolios formed each month on the return predictions from the three leading synthetic factors used jointly (KAPM3) or individually ($F_1$, $F_2$, $F_3$). Stocks are sorted into ten portfolios each month, and the high-minus-low (HML) strategy is the tenth minus the first decile return, marked "10−1." The sample covers 1987–2023 and excludes micro- and nano-cap stocks following the JKP size classification. Returns and volatilities are annualized and reported in percent. We report value-weighted (Panel A) and equal-weighted (Panel B) results.

| | KAPM3 | | | | KAPM ($F_1$) | | | | KAPM ($F_2$) | | | | KAPM ($F_3$) | | | |
|---|---|---|---|---|---|---|---|---|---|---|---|---|---|---|---|---|
| Decile | Ret | T-stat | Std | SR | Ret | T-stat | Std | SR | Ret | T-stat | Std | SR | Ret | T-stat | Std | SR |
| **Panel A: Value-Weighted** | | | | | | | | | | | | | | | | |
| 10 | 20.04 | 5.38 | 21.79 | 0.84 | 15.45 | 4.71 | 18.65 | 0.78 | 14.37 | 4.11 | 19.94 | 0.68 | 9.07 | 2.77 | 19.06 | 0.46 |
| 9 | 14.06 | 4.14 | 19.10 | 0.69 | 10.74 | 3.46 | 17.95 | 0.57 | 12.85 | 3.99 | 18.48 | 0.66 | 8.69 | 2.85 | 17.81 | 0.47 |
| 8 | 12.52 | 3.73 | 17.70 | 0.67 | 8.95 | 3.16 | 16.51 | 0.52 | 11.48 | 3.82 | 17.36 | 0.63 | 9.34 | 3.18 | 17.09 | 0.52 |
| 7 | 9.13 | 4.56 | 16.08 | 0.55 | 9.71 | 3.48 | 16.19 | 0.57 | 10.86 | 3.80 | 16.54 | 0.63 | 8.93 | 3.24 | 16.08 | 0.53 |
| 6 | 9.67 | 3.00 | 16.11 | 0.57 | 10.84 | 3.91 | 16.04 | 0.64 | 9.78 | 3.52 | 16.14 | 0.58 | 8.05 | 2.96 | 15.92 | 0.49 |
| 5 | 9.30 | 3.11 | 15.19 | 0.59 | 8.45 | 3.20 | 15.46 | 0.53 | 8.63 | 3.36 | 15.02 | 0.55 | 9.18 | 3.63 | 14.77 | 0.60 |
| 4 | 7.33 | 3.21 | 14.91 | 0.48 | 8.11 | 3.21 | 14.80 | 0.53 | 8.77 | 3.34 | 15.34 | 0.55 | 7.71 | 2.98 | 15.13 | 0.49 |
| 3 | 6.02 | 2.92 | 14.19 | 0.41 | 7.56 | 2.91 | 15.27 | 0.48 | 7.40 | 2.79 | 15.56 | 0.46 | 7.59 | 2.99 | 15.48 | 0.47 |
| 2 | 5.12 | 1.09 | 14.59 | 0.34 | 5.68 | 2.11 | 15.94 | 0.35 | 5.53 | 2.08 | 15.75 | 0.34 | 5.79 | 2.15 | 15.92 | 0.35 |
| 1 | -1.01 | -0.21 | 19.61 | -0.04 | 2.17 | 0.66 | 19.61 | 0.11 | -0.62 | -0.19 | 19.68 | -0.03 | 7.14 | 2.28 | 18.43 | 0.38 |
| 10−1 | 21.05 | 8.38 | 14.70 | 1.43 | 13.28 | 5.10 | 14.66 | 0.84 | 14.99 | 5.45 | 15.74 | 0.90 | 1.93 | 0.73 | 14.92 | 0.12 |
| FF6 α | 1.75 | 6.61 | - | - | 1.21 | 3.76 | - | - | 1.64 | 4.02 | - | - | 0.03 | 0.11 | - | - |
| **Panel B: Equal-Weighted** | | | | | | | | | | | | | | | | |
| 10 | 20.91 | 5.14 | 22.86 | 0.84 | 15.79 | 4.46 | 20.09 | 0.73 | 16.35 | 4.25 | 21.77 | 0.70 | 12.33 | 3.37 | 21.06 | 0.55 |
| 9 | 15.84 | 4.34 | 20.62 | 0.72 | 13.53 | 4.00 | 19.33 | 0.66 | 14.08 | 3.92 | 20.50 | 0.65 | 11.49 | 3.35 | 19.81 | 0.55 |
| 8 | 13.72 | 4.15 | 19.21 | 0.67 | 12.45 | 3.89 | 18.40 | 0.64 | 14.06 | 4.14 | 19.39 | 0.68 | 10.64 | 3.20 | 19.25 | 0.53 |
| 7 | 11.43 | 4.16 | 17.99 | 0.60 | 11.20 | 3.63 | 17.80 | 0.60 | 11.72 | 3.67 | 18.40 | 0.60 | 10.49 | 3.29 | 18.50 | 0.54 |
| 6 | 11.31 | 3.46 | 17.42 | 0.62 | 11.37 | 3.66 | 17.94 | 0.60 | 11.91 | 3.83 | 17.92 | 0.63 | 9.64 | 3.06 | 18.33 | 0.50 |
| 5 | 9.78 | 2.87 | 16.85 | 0.56 | 10.07 | 3.30 | 17.74 | 0.54 | 9.67 | 3.29 | 17.13 | 0.54 | 9.88 | 3.26 | 17.53 | 0.54 |
| 4 | 7.42 | 2.92 | 16.41 | 0.44 | 9.66 | 3.18 | 17.68 | 0.52 | 9.07 | 3.07 | 17.24 | 0.51 | 9.62 | 3.22 | 17.40 | 0.53 |
| 3 | 8.09 | 2.79 | 16.39 | 0.48 | 8.29 | 2.71 | 17.91 | 0.45 | 7.17 | 2.43 | 17.40 | 0.40 | 8.90 | 2.16 | 17.33 | 0.49 |
| 2 | 5.09 | 1.67 | 17.38 | 0.29 | 6.38 | 2.02 | 18.57 | 0.33 | 6.39 | 2.10 | 17.93 | 0.35 | 9.28 | 2.86 | 18.89 | 0.47 |
| 1 | -1.96 | -0.57 | 23.52 | -0.11 | 1.30 | 0.34 | 23.34 | 0.06 | -0.12 | -0.03 | 21.61 | -0.01 | 7.22 | 2.01 | 21.06 | 0.33 |
| 10−1 | 22.87 | 10.22 | 12.84 | 1.78 | 14.49 | 5.79 | 14.12 | 0.95 | 16.47 | 6.97 | 13.39 | 1.15 | 5.11 | 2.05 | 13.94 | 0.34 |
| FF6 α | 1.82 | 7.47 | - | - | 1.11 | 3.26 | - | - | 1.56 | 5.10 | - | - | 0.09 | 0.37 | - | - |

## Table 5: Drawdowns, Turnover, Higher Moments, and Risk-Adjusted Performance of Anomaly-Based and KAPM Portfolios

This table reports the maximum drawdown ("MaxDD"), the most extreme negative monthly return ("Max 1M Loss"), and the average monthly change in holdings ("Turnover"), all in percent, for the leading long–short portfolios. "KAPM8" is the full eight-factor portfolio; "KAPM3" uses the three leading synthetic factors jointly; and $F_1$, $F_2$, $F_3$ use each leading factor individually. Portfolio constructions follow Tables 1 and 4.

| | Anomalies | KAPM8 | KAPM3 | KAPM ($F_1$) | KAPM ($F_2$) | KAPM ($F_3$) |
|---|---|---|---|---|---|---|
| **Panel A: Value-Weighted** | | | | | | |
| MaxDD (%) | -11.51 | -30.35 | -28.06 | -46.22 | -29.71 | -57.24 |
| Max 1M Loss (%) | -5.55 | -12.95 | -12.52 | -14.51 | -10.88 | -22.93 |
| Turnover (%) | 26.03 | 69.93 | 65.32 | 64.57 | 65.70 | 65.62 |
| Skewness | 0.12 | 0.29 | 0.33 | 0.18 | 1.35 | -0.08 |
| Kurtosis | 3.49 | 1.49 | 2.48 | 1.77 | 6.53 | 5.41 |
| Info Ratio | 0.51 | 1.31 | 1.34 | 0.70 | 0.81 | 0.02 |
| **Panel B: Equal-Weighted** | | | | | | |
| MaxDD (%) | -8.11 | -27.07 | -19.16 | -49.52 | -30.46 | -55.41 |
| Max 1M Loss (%) | -4.37 | -11.75 | -9.93 | -17.30 | -13.21 | -20.98 |
| Turnover (%) | 26.35 | 60.65 | 57.49 | 60.17 | 61.49 | 61.82 |
| Skewness | -0.27 | 0.02 | 0.51 | -0.29 | 1.74 | -0.67 |
| Kurtosis | 2.66 | 1.57 | 2.16 | 2.55 | 14.39 | 5.51 |
| Info Ratio | 0.75 | 1.53 | 1.57 | 0.76 | 1.03 | 0.08 |

## Table 6: Cross-Sectional Tests: KAPM3 Factors

This table reports stock-level cross-sectional asset-pricing tests using the KAPM3 factors. The dependent variable is each stock's time-series average realized excess return; the explanatory variables are the time-series average contributions of the KAPM3 factors. Columns (1)–(5) progressively control for CAPM, FF3, FF5, and FF6 exposures. The FF6-only benchmark is reported in Table 2, column (5). Column 6 uses the KAPM signal produced by the three leading factors only instead of using all eight factors, which mirrors the one-signal cross-sectional testing setup of Table 2. Fama–MacBeth *t*-statistics are reported in parentheses. The standard error of a coefficient is the time-series standard deviation of its monthly estimates divided by the square root of the number of months. *, **, *** denote significance at the 10%, 5%, and 1% levels.

| | (1) Bare | (2) CAPM | (3) FF3 | (4) FF5 | (5) FF6 | (6) FF6 |
|---|---|---|---|---|---|---|
| $F_1$ | 1.701*** | 1.669*** | 1.720*** | 1.702*** | 1.681*** | |
| | (2.98) | (2.90) | (2.94) | (2.93) | (2.90) | |
| $F_2$ | 2.745* | 2.752* | 2.725* | 2.726* | 2.712* | |
| | (1.91) | (1.91) | (1.91) | (1.93) | (1.92) | |
| $F_3$ | -0.766* | -0.738* | -0.670* | -0.696* | -0.706* | |
| | (-1.84) | (-1.82) | (-1.75) | (-1.73) | (-1.83) | |
| KAPM3 | | | | | | 1.150*** |
| | | | | | | (2.93) |
| MKT−RF | | -0.121 | 0.071 | 0.091 | 0.063 | 0.040 |
| | | (-0.42) | (0.27) | (0.40) | (0.30) | (0.20) |
| SMB | | | -0.190 | -0.156 | -0.093 | -0.133 |
| | | | (-0.91) | (-0.83) | (-0.56) | (-0.78) |
| HML | | | -0.191 | -0.106 | -0.036 | -0.031 |
| | | | (-1.13) | (-0.71) | (-0.23) | (-0.20) |
| RMW | | | | 0.027 | 0.032 | 0.065 |
| | | | | (0.24) | (0.32) | (0.62) |
| CMA | | | | -0.043 | -0.039 | -0.043 |
| | | | | (-0.44) | (-0.43) | (-0.47) |
| MOM | | | | | 0.088 | 0.086 |
| | | | | | (0.58) | (0.57) |
| α | 0.135 | 0.289 | 0.217 | 0.156 | 0.141 | 0.048 |
| | (0.22) | (0.42) | (0.31) | (0.23) | (0.22) | (0.19) |
| Adj. $R^2$ | 0.052 | 0.053 | 0.064 | 0.058 | 0.054 | 0.028 |
| $R^2$ | 0.053 | 0.054 | 0.066 | 0.060 | 0.056 | 0.030 |
| Months | 443 | 443 | 443 | 443 | 443 | 443 |
| Stocks | 3910 | 3910 | 3910 | 3910 | 3910 | 3910 |

## Table 7: Cross-Sectional Tests: KAPM3 Synthetic Factors Across Benchmark Models

In this table, we report the alpha and beta estimates in cross-sectional asset-pricing tests of established factors and statistical factors, either including or excluding the three leading KAPM3 synthetic factors. The *Without KAPM3 α* is the intercept from the model without the KAPM3 factors; the *With KAPM3 α* and the KAPM3 loadings $F_1$–$F_3$ are from the specification that adds the three KAPM3 test factors. Established models (Panel A) each combine the Fama–French 3 factors (MKT, SMB, HML) with the model's own legs, as indicated by the row labels: the q-factor model (Hou–Xue–Zhang), Pastor–Stambaugh traded liquidity, Quality-Minus-Junk (AFP), Betting-Against-Beta (FP), and the Stambaugh–Yuan mispricing factors (MGMT, PERF). Statistical models (Panel B) use K = 3 factors each. Fama–MacBeth *t*-statistics are reported in parentheses. The standard error of a coefficient is the time-series standard deviation of its monthly estimates divided by the square root of the number of months. *, **, *** denote significance at the 10%, 5%, and 1% levels.

| | Without KAPM3 | With KAPM3 | | | |
|---|---|---|---|---|---|
| | (1) | (2) | (3) | (4) | (5) |
| | $\alpha$ | $\alpha$ | $F_1$ | $F_2$ | $F_3$ |
| *Panel A: Established Factor Models* | | | | | |
| FF3 + q-factor | 0.592*** | 0.107 | 1.767*** | 2.698* | -0.757* |
| | (2.62) | (0.16) | (3.04) | (1.81) | (-1.90) |
| FF3 + Liquidity | 0.641** | 0.179 | 1.747*** | 2.755* | -0.709* |
| | (2.51) | (0.26) | (3.04) | (1.85) | (-1.79) |
| FF3 + Quality | 0.566** | 0.094 | 1.690*** | 2.576* | -0.744* |
| | (2.30) | (0.14) | (2.90) | (1.72) | (-1.83) |
| FF3 + BAB | 0.523*** | 0.109 | 1.736*** | 2.747* | -0.768* |
| | (2.84) | (0.16) | (2.98) | (1.82) | (-1.70) |
| FF3 + Mispricing | 0.599*** | 0.115 | 1.738*** | 2.643* | -0.664* |
| | (3.35) | (0.17) | (3.01) | (1.76) | (-1.93) |
| *Panel B: Statistical Factor Models* | | | | | |
| PCA | 0.625** | 0.236 | 1.650*** | 2.778* | -0.749* |
| | (2.54) | (0.35) | (2.85) | (1.85) | (-1.75) |
| RP-PCA | 0.595** | 0.181 | 1.694*** | 2.763* | -0.745* |
| | (2.21) | (0.27) | (3.01) | (1.86) | (-1.75) |
| K-PCA | 0.889*** | 0.455 | 1.704*** | 2.743** | -0.780* |
| | (3.71) | (0.68) | (2.96) | (1.99) | (-1.74) |
| IPCA | 0.650** | 0.316 | 1.655*** | 2.845* | -0.774* |
| | (2.50) | (0.50) | (2.99) | (1.88) | (-1.72) |

## Table 8: Sources of Nonlinearity: Layer-Wise Ablation

This table isolates the contribution of nonlinearity at each layer of the KAPM models. The inner layer aggregates firm characteristics into synthetic factors; the outer layer prices those factors. Each column corresponds to a model in which the inner and outer layers are restricted to be linear ("L") or retained as nonlinear ("NL"), as indicated by the two header rows. The linear layers are obtained by applying post-hoc, globally linear approximation to the trained KAPM8 or KAPM3 models, and no re-training takes place. Because our KAN algorithm selects an order of three, the nonlinear relationships include constant, linear, quadratic and cubic ones. "Anomalies" repeats the anomaly benchmark of Table 1, and "PCA" is an out-of-sample long–short portfolio formed from the first three principal components of anomaly portfolio returns. Column (5) is equivalent to the full KAPM8 model in Table 1, nonlinear at both layers. Statistics are computed as in Table 1.

| | (1) | (2) | (3) | (4) | (5) | (6) | (7) | (8) |
|---|---|---|---|---|---|---|---|---|
| | Anomalies | PCA | | KAPM8 | | | KAPM3 | |
| Inner Layer | L | L | NL | L | NL | NL | L | NL |
| Outer Layer | L | L | L | NL | NL | L | NL | NL |
| *Panel A: Value-Weighted Portfolio* | | | | | | | | |
| Sharpe | 0.56 | 0.52 | 1.19 | 1.25 | 1.38 | 1.26 | 1.28 | 1.43 |
| Ret (%) | 2.39 | 9.20 | 16.31 | 16.83 | 20.19 | 17.04 | 17.89 | 21.05 |
| Vol (%) | 4.30 | 17.74 | 13.68 | 13.51 | 13.50 | 13.51 | 13.93 | 14.70 |
| MaxDD (%) | -11.51 | -58.27 | -29.15 | -29.72 | -30.35 | -24.66 | -21.20 | -28.06 |
| Skewness | 0.12 | 0.20 | 0.62 | 0.11 | 0.29 | 0.65 | 0.08 | 0.33 |
| Kurtosis | 3.49 | 1.74 | 2.90 | 1.74 | 1.49 | 3.21 | 1.57 | 2.48 |
| Alpha | 0.21 | 0.76 | 1.36 | 1.36 | 1.74 | 1.45 | 1.50 | 1.75 |
| Alpha t-stat | 2.82 | 2.45 | 5.49 | 5.63 | 6.01 | 5.64 | 6.39 | 6.61 |
| Info Ratio | 0.51 | 0.47 | 1.04 | 1.08 | 1.31 | 1.10 | 1.16 | 1.34 |
| *Panel B: Equal-Weighted Portfolio* | | | | | | | | |
| Sharpe | 0.83 | 0.69 | 1.47 | 1.55 | 1.68 | 1.54 | 1.60 | 1.78 |
| Ret (%) | 3.03 | 10.41 | 18.49 | 17.93 | 22.13 | 19.14 | 19.27 | 22.87 |
| Vol (%) | 3.65 | 14.99 | 12.54 | 11.50 | 12.21 | 12.44 | 12.08 | 12.84 |
| MaxDD (%) | -8.11 | -43.14 | -30.81 | -28.68 | -27.07 | -25.87 | -19.05 | -19.16 |
| Skewness | -0.27 | 0.07 | -0.02 | -0.08 | 0.02 | 0.11 | 0.40 | 0.51 |
| Kurtosis | 2.66 | 2.58 | 1.98 | 2.47 | 1.57 | 1.64 | 3.02 | 2.16 |
| Alpha | 0.25 | 0.89 | 1.52 | 1.51 | 1.78 | 1.58 | 1.65 | 1.82 |
| Alpha t-stat | 4.55 | 3.76 | 6.13 | 6.24 | 6.24 | 6.35 | 6.39 | 7.47 |
| Info Ratio | 0.75 | 0.63 | 1.28 | 1.37 | 1.53 | 1.33 | 1.41 | 1.57 |

## Table 9: Cross-Sectional Tests: Ablation "NL/L" Models

This table takes the ablated KAPM models in Table 8 and runs cross-sectional asset-pricing tests of their respective factors. We look at the "NL/L" models where the inner layer is nonlinear, and the outer layer is set to be linear. "KAPM8" refers to the KAPM8 model exhibited in Column (3) of Table 8, and "KAPM3" refers to the KAPM3 model exhibited in Column (6) of Table 8. The factors refer to the leading three factors of KAPM3. The dependent variable is each stock's time-series average realized excess return; the explanatory variables are the time-series average values of the model-implied conditional expected returns of the KAPM8 model, the KAPM3 model or contributions of the KAPM3 factors. Columns (4)–(6) control for FF6 exposures. Fama–MacBeth *t*-statistics are reported in parentheses. The standard error of a coefficient is the time-series standard deviation of its monthly estimates divided by the square root of the number of months. *, **, *** denote significance at the 10%, 5%, and 1% levels.

| | (1)<br>Bare | (2)<br>Bare | (3)<br>Bare | (4)<br>FF6 | (5)<br>FF6 | (6)<br>FF6 |
|---|---|---|---|---|---|---|
| KAPM8 | 0.456** | | | 0.553** | | |
| | (2.18) | | | (2.76) | | |
| KAPM3 | | 0.835*** | | | 0.837*** | |
| | | (3.85) | | | (4.33) | |
| $F_1$ | | | 2.161** | | | 2.238** |
| | | | (2.30) | | | (2.52) |
| $F_2$ | | | 0.994* | | | 1.024* |
| | | | (1.80) | | | (1.87) |
| $F_3$ | | | -1.838** | | | -1.864** |
| | | | (-2.29) | | | (-2.35) |
| MKT−RF | | | | 0.166 | 0.057 | 0.149 |
| | | | | (0.73) | (0.28) | (0.76) |
| SMB | | | | -0.239 | -0.091 | -0.054 |
| | | | | (-0.64) | (-0.54) | (-0.34) |
| HML | | | | -0.132* | -0.068 | -0.097 |
| | | | | (-1.69) | (-0.45) | (-0.64) |
| RMW | | | | -0.003 | 0.042 | 0.008 |
| | | | | (-0.03) | (0.41) | (0.08) |
| CMA | | | | -0.105 | -0.068 | -0.084 |
| | | | | (-0.44) | (-0.75) | (-0.94) |
| MOM | | | | 0.149 | 0.095 | 0.115 |
| | | | | (1.56) | (0.63) | (0.76) |
| α | 0.346** | 0.014 | -0.021 | 0.362** | 0.030 | -0.136 |
| | (2.18) | (0.04) | (-0.06) | (2.49) | (0.14) | (-0.57) |
| Adj. $R^2$ | 0.020 | 0.034 | 0.084 | 0.024 | 0.037 | 0.089 |
| $R^2$ | 0.021 | 0.036 | 0.085 | 0.026 | 0.039 | 0.091 |
| Months | 443 | 443 | 443 | 443 | 443 | 443 |
| Stocks | 3910 | 3910 | 3910 | 3910 | 3910 | 3910 |

## Figure 1: Cumulative Returns of the KAPM and Average-Anomaly Portfolios

This figure reports the cumulative compounded returns of three equally weighted long–short portfolios over 1987–2023. Each portfolio sorts stocks into ten deciles on a one-month-ahead out-of-sample signal and buys the top decile while selling the bottom decile, compounding returns through time. The red line is the average anomaly portfolio, the cross-anomaly average of the 152 individual high-minus-low portfolios, identical to the equally weighted anomaly benchmark in Table 1. The black line is the KAPM portfolio formed from the model's overall (eight-factor) return predictions, and the blue line is the KAPM portfolio formed from the predictions of the three leading factors. Because the KAPM portfolios accumulate far larger cumulative returns than the average anomaly portfolio, they are plotted on a separate scale: the left axis ("Anomalies Portfolio") applies to the red line, and the right axis ("KAPM Portfolios") applies to the black and blue lines. The terminal cumulative value of each portfolio is annotated at the right edge.

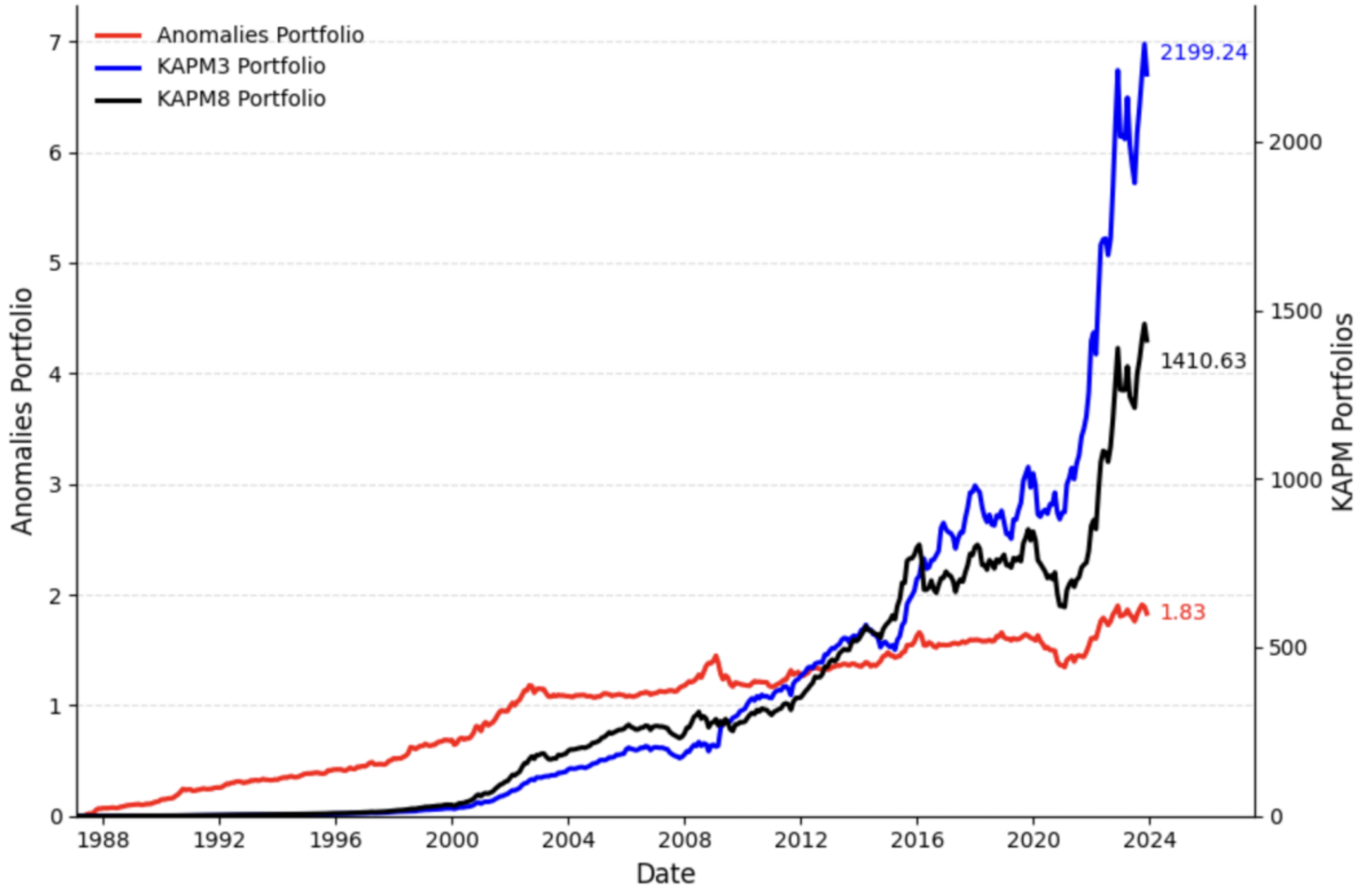

## Figure 2: Out-of-Sample Sharpe Ratios of Ranked Synthetic-Factor Portfolios

This figure plots the out-of-sample annualized Sharpe ratio of the equally weighted long–short portfolio formed on the return predictions of the $K$ leading synthetic factors, for $K = 1, \ldots, 8$ (horizontal axis). The factors are ranked by the importance score

$$S_q = \frac{I_q^{in}}{\sum_{j=1}^{8} I_j^{in}} \cdot \frac{I_q^{out}}{\sum_{j=1}^{8} I_j^{out}},$$

where $I_q^{in} = \sum_p \sum_m \left|\theta_{q,p,m}^{\phi}\right|$ and $I_q^{out} = \sum_m \left|\theta_{q,m}^{\Phi}\right|$ are the sums of the absolute values of the inner-layer and outer-layer spline coefficients of factor $q$. Because the ranking depends only on ex-ante trained weights, the sorting is immune to look-ahead bias. Each month, stocks are sorted into deciles on the return prediction based on the $K$ leading factors, and the portfolio buys the top decile and sells the bottom decile. The sample period is 1987–2023. The dotted line marks the Sharpe ratio of the full eight-factor specification (1.68); the maximum, an annualized 1.78, is reached by the three-factor portfolio.

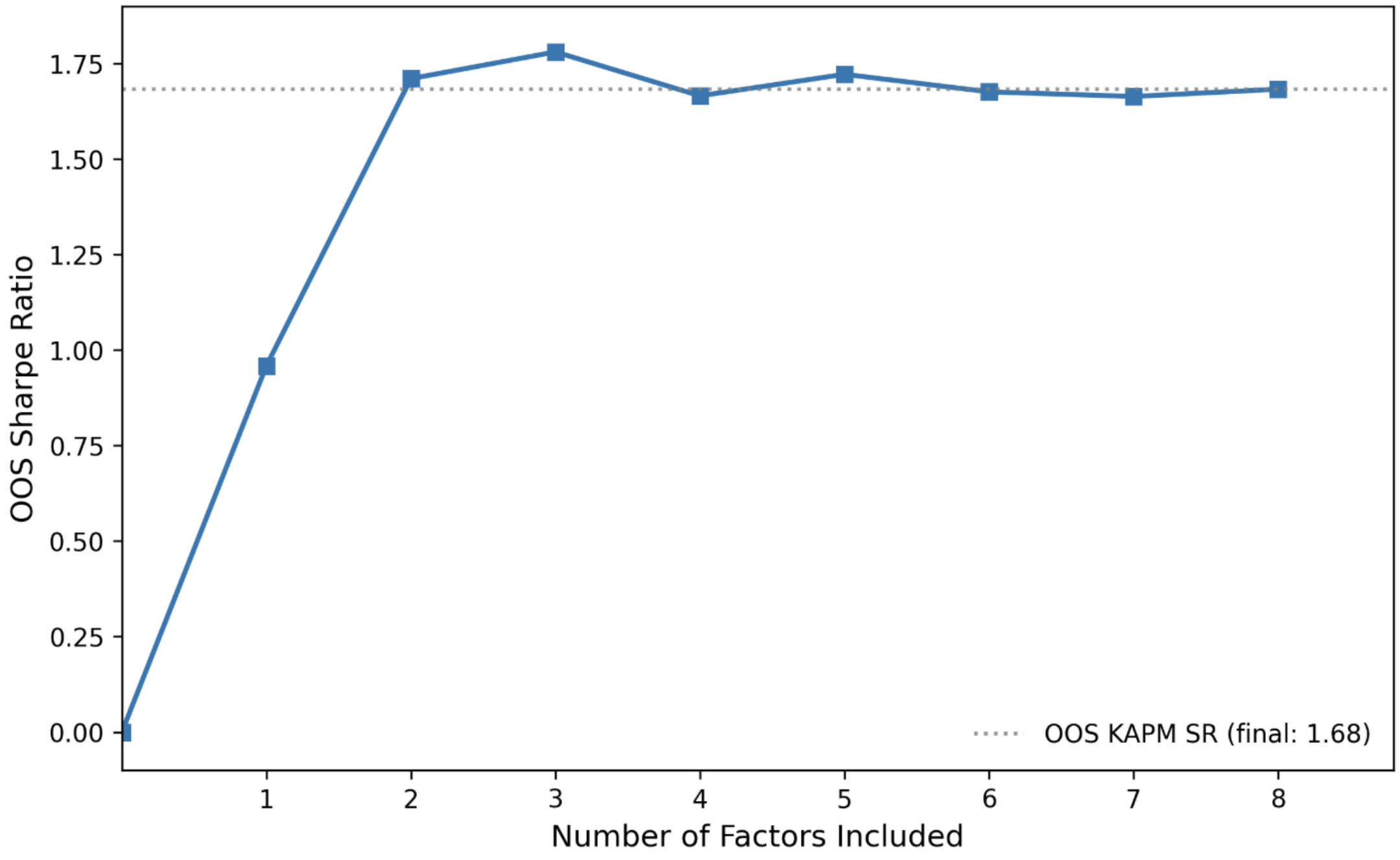

## Figure 3: Recursive Bayesian Factor Selection over Time

This figure reports recursive Bayesian factor-selection results for the eight KAPM8 synthetic factors, F1–F8. The first 44 months (January 1987–August 1990) are reserved as a fixed training period to construct the prior mean of expected factor returns. Bayesian model comparison then uses an expanding estimation window that begins in September 1990; the first 36 months of this window (September 1990–August 1993) serve as a burn-in period, so results are plotted from August 1993 through November 2023. At each recursive date we evaluate all 255 non-empty subsets of the eight factors using CZZ Proposition 5. Panel A reports the recursive posterior inclusion probability of each factor; Panel B plots the posterior probabilities of the seven models with the largest end-of-sample posterior probability.

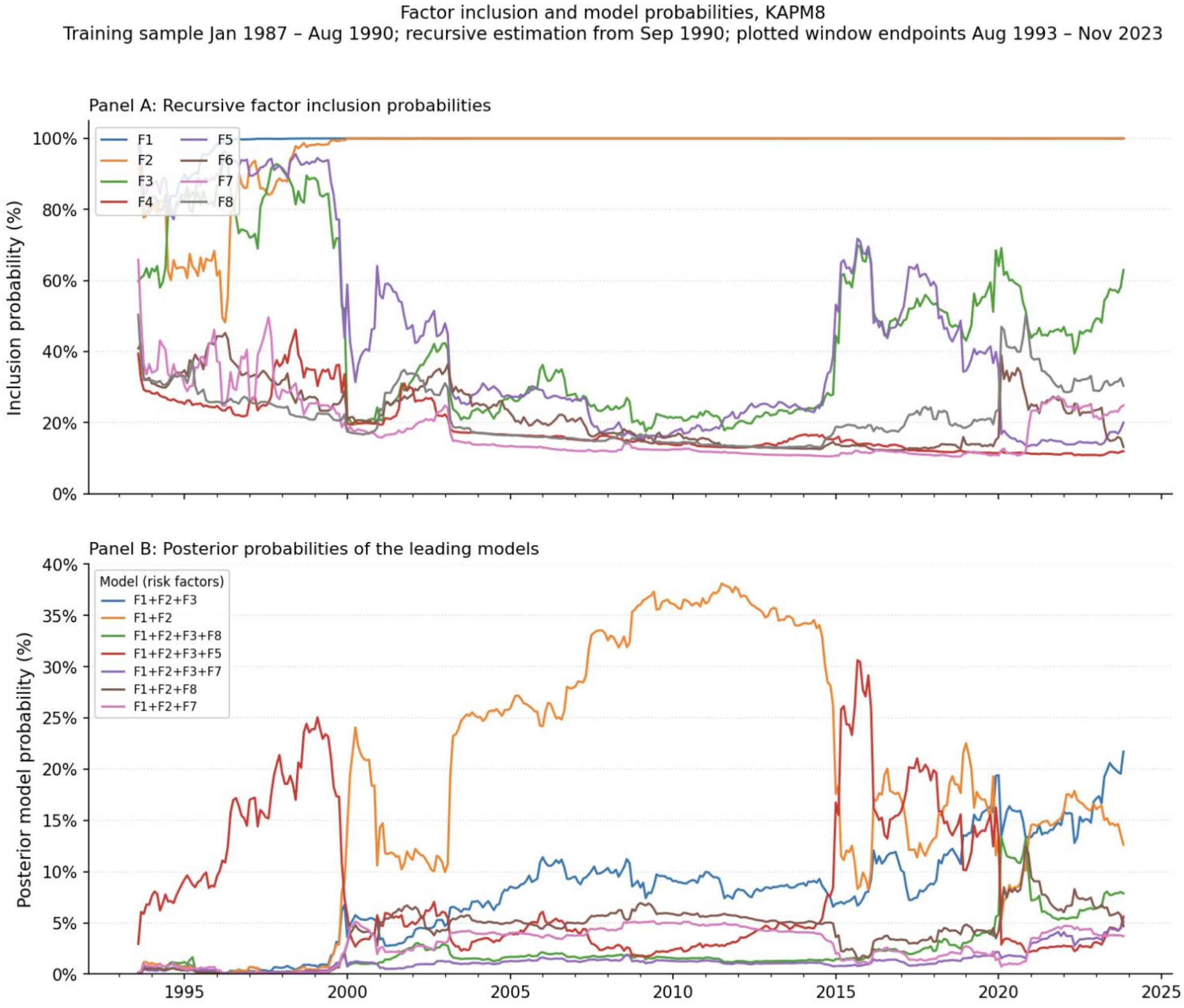

## Figure 4: KAPM3 Factors versus the Anomaly Zoo in the Cross-Section

This figure compares the cross-sectional explanatory power of the 152 anomaly factors in the JKP library with that of the KAPM3 factors. For each anomaly factor, we estimate stock-level exposures from time-series regressions and then run cross-sectional regressions of average stock returns on those exposures. The left panel reports the anomaly-only specification $E[r_i] = \alpha + \lambda^{zoo,j}\beta_i^{zoo,j} + \varepsilon_i$; the right panel augments it with the three leading KAPM3 factors, $E[r_i] = \alpha + \sum_{q=1}^{3}\gamma_q\, E[\Phi_{q,i}] + \lambda^{zoo,j}\beta_i^{zoo,j} + \varepsilon_i$. The *t*-statistics are based on Fama–MacBeth standard errors. The standard error of a coefficient is the time-series standard deviation of its monthly estimates divided by the square root of the number of months. Each panel reports the average absolute t-statistics across the 152 cross-sectional regressions; error bars are one standard deviation across anomalies, and the dashed line marks the 10% significance threshold (|t| = 1.64).

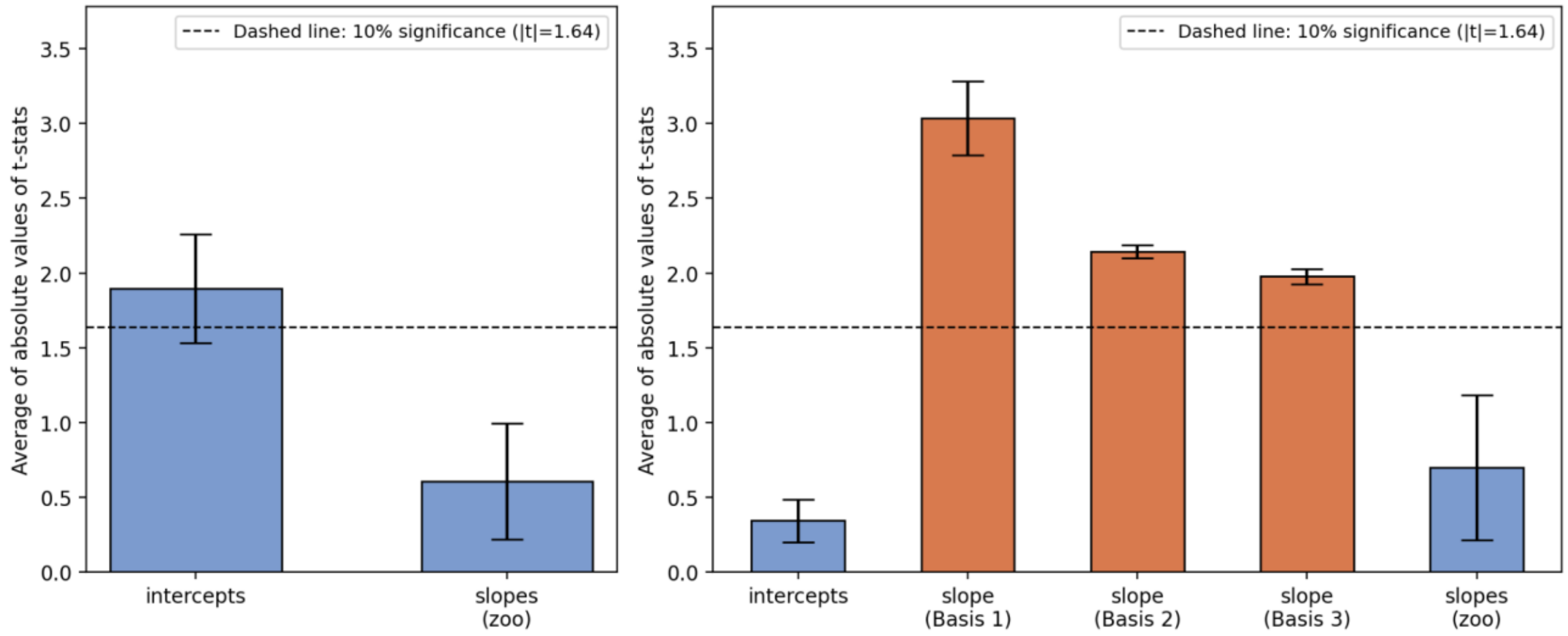

| | Left Figure | | Right Figure | | | | |
|---|---|---|---|---|---|---|---|
| | Intercept | Slope (zoo) | Intercept | Slope ($F_1$) | Slope ($F_2$) | Slope ($F_3$) | Slope (zoo) |
| Mean | 1.89 | 0.61 | 0.34 | 3.04 | 2.14 | 1.98 | 0.70 |
| 1% | 1.39 | 0.02 | 0.01 | 2.73 | 2.05 | 1.88 | 0.02 |
| 5% | 1.44 | 0.13 | 0.11 | 2.76 | 2.08 | 1.92 | 0.06 |
| 10% | 1.62 | 0.17 | 0.18 | 2.80 | 2.09 | 1.93 | 0.10 |
| 25% | 1.64 | 0.30 | 0.26 | 2.90 | 2.12 | 1.95 | 0.29 |
| 50% | 1.75 | 0.52 | 0.32 | 2.97 | 2.14 | 1.96 | 0.67 |
| 75% | 2.00 | 0.84 | 0.43 | 3.05 | 2.17 | 1.99 | 1.07 |
| 90% | 2.48 | 1.17 | 0.56 | 3.41 | 2.19 | 2.04 | 1.39 |
| 95% | 2.73 | 1.30 | 0.60 | 3.57 | 2.21 | 2.07 | 1.51 |
| 99% | 3.07 | 1.58 | 0.67 | 3.84 | 2.28 | 2.14 | 1.89 |

## Figure 5: Theme Composition of the KAPM3 Factors

This figure reports the relative importance of economic themes for each of the KAPM3 factors (Factor 1–3, from left to right). For each of the 760 model inputs and each factor, we compute an importance score as the sum of the absolute values of the corresponding inner-layer spline coefficients; we sum the scores of an anomaly's base term and its four macroeconomic interactions into a single anomaly-level score, group the 152 anomalies into the thirteen themes of the JKP taxonomy (value, momentum, low risk, and so on), and take the per-anomaly (count-weighted) average within each theme. Normalizing across themes so that the shares for a given factor sum to one, the relative importance of theme $g$ for factor $q$ is $\omega_{q,g} = \frac{(1/N_g)\sum_{a\in g} I_{q,a}}{\sum_{g'}(1/N_{g'})\sum_{a\in g'} I_{q,a}}$, where $N_g$ is the number of anomalies in theme $g$ and $I_{q,a}$ is the anomaly-level importance of anomaly $a$ for factor $q$. Within each panel, themes are ordered from most to least important (top to bottom).

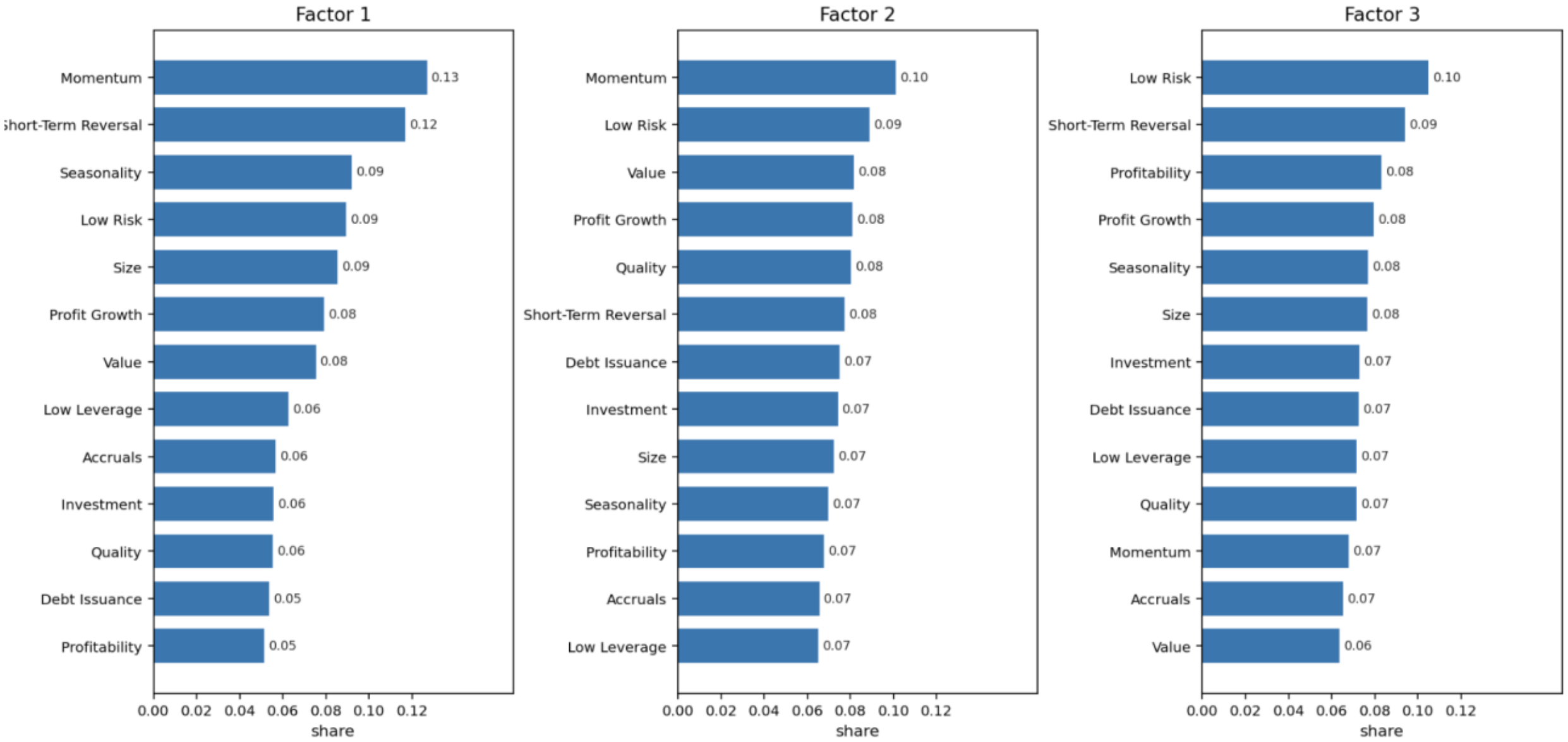

# Internet Appendix

## The Hidden Structure of the Factor Zoo: When Asset Pricing Meets Kolmogorov–Arnold

This Internet Appendix provides supplementary material for the paper. Part I contains supplementary discussions and methods. Section 1 describes the estimation and implementation of KAN, including the B-spline representation, the two-layer network parameterization, the loss function, optimization, and the training algorithm. Section 2 documents the architecture-selection analysis. Section 3 describes the Bayesian selection of synthetic factors. Section 4 reports supplementary empirical results, including a benchmark against the neural-network results of Gu, Kelly, and Xiu (2020) and additional cross-sectional pricing tests. Section 5 describes the recovery and interpretation of the synthetic-factor functions. Part II contains the supplementary tables and figures.

# Part I. Supplementary Discussions and Methods

## 1 KAN Estimation and Implementation Details

### *1.1 B-Spline Representation*

This section describes the numerical implementation of KAN and the estimation procedure used in the main paper.

Following Liu et al. (2025), we approximate learnable univariate functions with B-splines. For a generic function $g(x)$, the spline approximation is

$$g(x) = \sum_{m=1}^{M} \theta_m \, B_m^{(d)}(x), \tag{IA1}$$

where $B_m^{(d)}(x)$ denotes the $m$-th B-spline basis function of polynomial degree $d$, $\theta_m$ is the corresponding coefficient, and $M$ is the number of basis functions.

Let the input domain $[a, b]$ be partitioned into $G$ intervals by the grid points

$$a = \xi_0 < \xi_1 < \cdots < \xi_G = b. \tag{IA2}$$

The grid is extended at the boundaries to construct the spline basis. Under the extended-grid convention used here, each edge function has $M = G + d$ spline coefficients.

The degree-zero basis functions are

$$B_m^{(0)}(x) = \begin{cases} 1, & \xi_m \le x < \xi_{m+1}, \\ 0, & \text{otherwise,} \end{cases} \tag{IA3}$$

and higher-degree functions are obtained recursively using the Cox–de Boor formula,

$$B_m^{(d)}(x) = \frac{x - \xi_m}{\xi_{m+d} - \xi_m} B_m^{(d-1)}(x) + \frac{\xi_{m+d+1} - x}{\xi_{m+d+1} - \xi_{m+1}} B_{m+1}^{(d-1)}(x). \tag{IA4}$$

Terms with zero denominators are defined to be zero. Cubic splines correspond to $d = 3$, yielding piecewise cubic functions that are smooth at ordinary interior knots.

The grid size and polynomial degree determine the flexibility of the estimated functions. A finer grid allows more localized nonlinear variation but increases the number of parameters. A higher polynomial degree changes the smoothness and complexity of the approximation. Our preferred specification uses cubic splines.

### *1.2 Two-Layer KAN Parameterization*

Applying the spline approximation to the inner-layer functions gives

$$\phi_{q,p}\big(Z_{i,p,t}\big) = \sum_{m=1}^{M} \theta_{q,p,m}^{\phi} \, B_{q,p,m}^{\phi,(d)}\big(Z_{i,p,t}\big), \qquad p = 1, \dots, L,\ q = 1, \dots, Q, \tag{IA5}$$

where each input-to-factor connection has its own learnable coefficients and corresponding spline basis. The synthetic factors are constructed by aggregating these transformations:

$$s_{i,q,t} = \sum_{p=1}^{L} \phi_{q,p}\left(Z_{i,p,t}\right). \tag{IA6}$$

Similarly, the outer-layer functions are parameterized as

$$\Phi_q\left(s_{i,q,t}\right) = \sum_{m=1}^{M} \theta_{q,m}^{\Phi} B_{q,m}^{\Phi,(d)}\left(s_{i,q,t}\right). \tag{IA7}$$

Combining the two layers, the predicted expected excess return is

$$\hat{E}_t\left[r_{i,t+1}\right] = \sum_{q=1}^{Q} \Phi_q\left(s_{i,q,t}\right) = \sum_{q=1}^{Q} \Phi_q\left(\sum_{p=1}^{L} \phi_{q,p}\left(Z_{i,p,t}\right)\right). \tag{IA8}$$

This specification contains $QL$ input-to-factor functions and $Q$ factor-to-output functions, which the network estimates jointly from the training data. In our empirical specification, $\mathbf{Z}_{i,t}$ contains $L = 760$ inputs, consisting of firm characteristics and their interactions with macroeconomic state variables (Section III of the main paper).

The additive inner-layer representation also allows constant contributions common across assets to be incorporated into the learned transformations. Such contributions are not separately identified from the spline functions without additional restrictions. Our analysis focuses on the resulting characteristic-to-return mapping and its synthetic-factor structure.

### *1.3 Loss Function and Regularization*

Let $\mathbf{\Theta}$ collect the learnable network coefficients. We estimate these parameters by minimizing the regularized prediction loss,

$$\mathcal{L}(\mathbf{\Theta}) = \frac{1}{N_{\text{train}}} \sum_{(i,t)\in\mathcal{T}_{\text{train}}} \left(r_{i,t+1} - \hat{E}_t\left[r_{i,t+1}; \mathbf{\Theta}\right]\right)^2 + \eta\, \Omega(\mathbf{\Theta}), \tag{IA9}$$

where $\mathcal{T}_{\text{train}}$ denotes the training sample and $N_{\text{train}}$ is its number of stock-month observations. The regularization penalty is

$$\Omega(\mathbf{\Theta}) = \sum_{\theta\in\mathbf{\Theta}} |\theta|. \tag{IA10}$$

The $L_1$ penalty shrinks coefficients toward zero and reduces the complexity of the estimated nonlinear functions. The parameter $\eta$ governs the degree of shrinkage.

### *1.4 Optimization and Anchoring*

We minimize the objective in equation (IA9) using AdamW, a gradient-based optimizer with decoupled weight decay. At each training iteration, the network evaluates the spline functions, aggregates the inner-layer transformations into synthetic factors, computes predicted returns, and updates the learnable parameters using the gradient of the loss.

Our empirical procedure uses expanding estimation windows. The initial training sample spans 1957–1986, and out-of-sample return predictions begin in 1987. The model is re-estimated annually as additional observations become available.

To improve computational efficiency and training stability, we use an anchoring procedure. Before estimating the model over the full expanding window, we first train a KAN using the preceding ten years of observations. Its estimated parameters are then used to initialize optimization over the full expanding window. The anchoring sample is contained in the estimation window and therefore does not introduce observations from the subsequent prediction period.

Within each estimation window, the learned spline coefficients are common across all training observations. The network is re-estimated as the sample window expands.

### *1.5 Validation and Hyperparameter Selection*

Similar to Gu, Kelly, and Xiu (2020), we divide each estimation sample into training and validation subsets, with proportions of 80% and 20%, respectively. For candidate estimation configuration $c$, the validation prediction loss is

$$\mathcal{L}_{\text{val}}(c) = \frac{1}{N_{\text{val}}} \sum_{(i,t)\in\mathcal{T}_{\text{val}}} \left(r_{i,t+1} - \hat{E}_t^{(c)}[r_{i,t+1}]\right)^2. \tag{IA11}$$

The training subset is used to fit the spline coefficients, while the validation subset evaluates predictive accuracy under alternative estimation settings. Grid size and regularization strength, which control function flexibility and shrinkage, are tuned in this way. The number of synthetic factors, spline degree, and network depth are selected using only the initial training sample (1957–1986), as described in Section 2.

### *1.6 Training Algorithm*

**Algorithm IA1: KAN Estimation and Validation**

**Input:** Training observations $\{(\mathbf{Z}_{i,t}, r_{i,t+1})\}$, spline grid size $G$, degree $d$, network architecture, regularization parameter $\eta$, and learning rate $\nu$.

**Output:** Estimated network parameters $\widehat{\boldsymbol{\Theta}}$ and return forecasts.

1. Normalize all characteristics to $[-1,1]$, and construct the training and validation subsets from the estimation window.
2. Construct the spline grids and B-spline basis functions using equations (IA2)–(IA4).
3. Initialize the network parameters, using the anchoring sample when applicable.
4. Evaluate the inner-layer spline functions and aggregate them into synthetic factors, equations (IA5)–(IA6).
5. Apply the outer-layer functions to obtain predicted excess returns, equations (IA7)–(IA8).
6. Calculate the penalized loss in equation (IA9) and update the parameters using AdamW. Repeat until the specified optimization procedure is completed.
7. Calculate the validation loss in equation (IA11) and assess candidate estimation configurations.
8. Use the estimated network to forecast returns over the subsequent prediction period, re-estimating annually with the expanding window.

## 2 Architecture Selection

This section examines how the estimated KAPM depends on three architectural choices: the number of synthetic factors, spline degree, and network depth. The number of synthetic factors determines the dimensionality of the estimated factor representation, so architecture selection addresses not only predictive performance but also whether the return-relevant information in a large set of characteristics can be summarized by a low-dimensional structure. We select the architecture using only the initial training sample (1957–1986), so that the out-of-sample evaluation over 1987–2023 is not affected by the selection.

For each candidate architecture, we construct high-minus-low portfolios based on fitted expected returns and calculate their Sharpe ratios. Figure IA1 reports these comparisons for the initial training sample, on which the selection is based. For comparison, it also reports the same statistics for the largest expanding training sample (1957–2022), which show that the ranking of architectures is stable as the sample expands. The architecture with one hidden layer and eight synthetic factors, denoted [8], performs particularly well.

We further assess the incremental information contained in portfolios generated by alternative architectures using Bayesian factor selection. Each architecture produces a portfolio return series, which is evaluated against those generated by competing specifications. A higher posterior inclusion probability indicates that the portfolio contributes information not readily captured by the others. Figure IA1 shows that the [8] architecture also receives strong support from this analysis.

Table IA1 reports out-of-sample performance under alternative specifications as a robustness check. Panel A varies spline degree and the number of synthetic factors in a one-hidden-layer KAN. Panel B compares alternative network depths and widths, holding the spline degree fixed at three. The results favor cubic splines and a relatively shallow network. Increasing the number of synthetic factors does not monotonically improve portfolio performance, and additional hidden layers generally do not produce corresponding gains.

The [8] specification produces annualized out-of-sample Sharpe ratios of 1.38 and 1.68 for value-weighted and equally weighted high-minus-low portfolios, respectively. Nearby architectures generate economically similar results, suggesting that performance is not highly sensitive to modest changes in network width. Overall, the evidence supports a parsimonious synthetic-factor representation of the characteristic-to-return mapping.

## 3 Bayesian Selection of Synthetic Factors

This section describes the Bayesian factor-selection procedure used in Section V.A of the main paper. We follow the model-comparison framework of Barillas and Shanken (2018), using the marginal likelihood approach of Chib, Zeng, and Zhao (2020, Section III.C) without prespecifying any synthetic factor as a risk factor.

Let $\mathbf{F}_t = (F_{1,t}, \dots, F_{Q,t})'$ denote the returns of the $Q = 8$ ranked synthetic-factor portfolios. Each candidate model $M_j$ divides these factors into a nonempty set of $h_j$ candidate pricing factors, $\mathbf{F}_{j,t}$, and the remaining $Q - h_j$ factors, $\mathbf{F}_{j,t}^*$. We consider all $2^Q - 1 = 255$ nonempty subsets. Under model $M_j$, factor returns satisfy

$$\mathbf{F}_{j,t} = \boldsymbol{\alpha}_j + \boldsymbol{\varepsilon}_{j,t}, \qquad \boldsymbol{\varepsilon}_{j,t} \sim N_{h_j}(\mathbf{0}, \boldsymbol{\Sigma}_j), \tag{IA12}$$

$$\mathbf{F}_{j,t}^* = \mathbf{B}_j^* \mathbf{F}_{j,t} + \boldsymbol{\varepsilon}_{j,t}^*, \qquad \boldsymbol{\varepsilon}_{j,t}^* \sim N_{Q-h_j}(\mathbf{0}, \boldsymbol{\Sigma}_j^*). \tag{IA13}$$

The zero-intercept restriction in equation (IA13) requires the remaining factors to be priced by the selected factors. Bayesian model comparison therefore evaluates alternative factor-spanning restrictions, allowing for uncertainty about which factors are needed to explain the expected returns of the others.

Following Chib, Zeng, and Zhao (2020, Proposition 4), the nuisance parameters receive the improper prior $\psi(\boldsymbol{\eta}_j \mid M_j) = |\boldsymbol{\Sigma}_j|^{-(2h_j - Q + 1)/2} |\boldsymbol{\Sigma}_j^*|^{-(Q+1)/2}$, which makes marginal likelihoods comparable across models. The factor means receive the model-specific prior $\boldsymbol{\alpha}_j \mid M_j \sim N_{h_j}(\boldsymbol{\alpha}_{j0}, k_j \boldsymbol{\Sigma}_j)$, where the prior mean is the average of the selected factors over a training sample of the first $n_t$ observations:

$$\boldsymbol{\alpha}_{j0} = \frac{1}{n_t} \sum_{t=1}^{n_t} \mathbf{F}_{j,t}. \tag{IA14}$$

We set $n_t = 44$, so that the training sample covers January 1987 through August 1990, approximately the first tenth of the 443-month evaluation period, as in Chib, Zeng, and Zhao (2020). The multiplier $k_j$ follows their equation (26). The remaining $\tilde{T} = T - n_t$ observations are used for model comparison. In the analysis of Figure 3, the estimation window expands from September 1990, with the first 36 months serving as an estimation burn-in period. Under these priors, the log marginal likelihood of $M_j$ separates into two closed-form terms (Chib, Zeng, and Zhao 2020, Proposition 5):

$$\log m(\mathbf{F}_{n_t+1:T} \mid M_j) = \log m(\mathbf{F}_j \mid M_j) + \log m(\mathbf{F}_j^* \mid M_j), \tag{IA15}$$

where

$$\log m(\mathbf{F}_j \mid M_j) = -\frac{(Q - h_j)h_j}{2}\log 2 - \frac{\tilde{T} h_j}{2}\log \pi - \frac{h_j}{2}\log(\tilde{T} k_j + 1) - \frac{\tilde{T} + h_j - Q}{2}\log|\boldsymbol{\Psi}_j| + \log \Gamma_{h_j}\left(\frac{\tilde{T} + h_j - Q}{2}\right),$$

$$\log m(\mathbf{F}_j^* \mid M_j) = \frac{(Q - h_j)h_j}{2}\log 2 - \frac{(Q - h_j)(\tilde{T} - h_j)}{2}\log \pi - \frac{Q - h_j}{2}\log|\mathbf{W}_j| - \frac{\tilde{T}}{2}\log|\boldsymbol{\Psi}_j^*| + \log \Gamma_{Q-h_j}\left(\frac{\tilde{T}}{2}\right).$$

Here, $\Gamma_d(\cdot)$ is the $d$-dimensional multivariate gamma function, hats denote least-squares estimates over the estimation sample, and

$$\boldsymbol{\Psi}_j = \sum_{t=n_t+1}^{T} \left(\mathbf{F}_{j,t} - \widehat{\boldsymbol{\alpha}}_j\right)\left(\mathbf{F}_{j,t} - \widehat{\boldsymbol{\alpha}}_j\right)' + \frac{\tilde{T}}{\tilde{T}k_j + 1}\left(\widehat{\boldsymbol{\alpha}}_j - \boldsymbol{\alpha}_{j0}\right)\left(\widehat{\boldsymbol{\alpha}}_j - \boldsymbol{\alpha}_{j0}\right)',$$

$$\mathbf{W}_j = \sum_{t=n_t+1}^{T} \mathbf{F}_{j,t}\,\mathbf{F}_{j,t}{}', \qquad \boldsymbol{\Psi}_j^* = \sum_{t=n_t+1}^{T} \left(\mathbf{F}_{j,t}^* - \widehat{\mathbf{B}}_j^*\mathbf{F}_{j,t}\right)\left(\mathbf{F}_{j,t}^* - \widehat{\mathbf{B}}_j^*\mathbf{F}_{j,t}\right)'.$$

We assign equal prior probabilities to all candidate models. The posterior probability of model $M_j$ and the posterior inclusion probability of factor $F_q$ are

$$P\left(M_j \mid D\right) = \frac{m\left(\mathbf{F}_{n_t+1:T} \mid M_j\right)}{\sum_{l=1}^{255} m\left(\mathbf{F}_{n_t+1:T} \mid M_l\right)}, \tag{IA16}$$

$$P\left(F_q \mid D\right) = \sum_{j:\,F_q \in \mathbf{F}_j} P\left(M_j \mid D\right). \tag{IA17}$$

Table 3 of the main paper reports these quantities for the full evaluation period. For the recursive analysis in Figure 3, we hold the training sample fixed and update the posterior probabilities as the estimation window expands. This exercise describes how support for the leading factors evolves as additional observations become available; it does not impose a three-factor specification at each recursive date.

## 4 Supplementary Empirical Results

### *4.1 Benchmark Against Gu, Kelly, and Xiu (2020)*

This section benchmarks KAPM8 against the neural-network results of Gu, Kelly, and Xiu (2020). Table IA2 compares portfolio performance over the full 1987–2023 sample and the overlapping 1987–2016 period.

Over the overlapping period, the value-weighted and equally weighted KAPM8 portfolios excluding microcaps generate annualized Sharpe ratios of 1.50 and 1.98, respectively. The corresponding neural-network portfolios reported by Gu, Kelly, and Xiu (2020), which include all stocks, have Sharpe ratios of 1.35 and 1.68. The table also reports portfolio returns, volatility, drawdowns, and risk-adjusted performance measures.

Although the comparison uses an overlapping sample period, the two studies differ in their predictor sets, estimation procedures, stock universes, and portfolio construction. The results therefore provide a benchmark against existing machine-learning asset-pricing evidence rather than a controlled comparison of the two architectures.

### *4.2 Additional Cross-Sectional Pricing Tests*

This subsection reports the full stock-level cross-sectional regressions summarized in Table 7 of the main paper. Table IA3 considers statistical factor models estimated with three components: PCA, RP-PCA, kernel PCA, and IPCA. Table IA4 considers established factor models, each augmented with the Fama–French three factors: the q-factor model of Hou, Xue, and Zhang (2015), the traded liquidity factor of Pástor and Stambaugh (2003), quality-minus-junk (Asness, Frazzini, and Pedersen 2019), betting-against-beta (Frazzini and Pedersen 2014), and the mispricing factors of Stambaugh and Yuan (2017). In each table, Panel A estimates the regression with the benchmark factor exposures alone, and Panel B adds the contributions of the three leading synthetic factors.

# 5 Recovery and Interpretation of Synthetic-Factor Functions

## *5.1 Outer-Layer Function Recovery*

This section describes how we recover and visualize the nonlinear transformations associated with the estimated synthetic factors. Each outer-layer function $\Phi_q(\cdot)$ is represented by the cubic B-spline expansion in equation (IA7), with basis functions constructed as described in Section 1.1. Conditional on the spline grid, the estimated coefficients determine the nonlinear mapping from each synthetic-factor activation to its contribution to predicted returns.

To visualize these functions, we evaluate each recovered transformation over the empirical support of the corresponding hidden-node activation. Specifically, for each selected factor, we construct a grid between the 0.5th and 99.5th percentiles of observed activations and evaluate the estimated spline function at each grid point.

We also approximate the recovered functions using global third-degree polynomials over several ranges of their empirical support. These polynomial approximations provide compact descriptions of the functions learned by KAN, but they are not imposed during estimation.

Table IA6 reports the approximations for the three leading synthetic factors, using the central 0.5–99.5%, 1–99%, 2.5–97.5%, and 5–95% support ranges. The reported reconstruction accuracy exceeds 99.5%. Figure IA2 illustrates the corresponding recovered nonlinear transformations.

# Part II. Supplementary Tables and Figures

## Table IA1: Out-of-Sample Sharpe Ratio of KAPM Portfolios: Spline Order, Factor Count, and Depth

This table reports the annualized out-of-sample Sharpe ratio of KAPM long–short portfolios under alternative architectures. Panel A varies the spline order (rows) and the number of latent synthetic factors (columns) within a one-layer KAN; architectures are labeled by the number of latent factors in brackets. Panel B varies the depth and width of the network while holding the spline order fixed at three, with each architecture labeled by its sequence of hidden-layer widths. VW and EW denote value- and equally weighted portfolios. The selected specification—a one-layer network with eight latent factors and cubic (order-3) splines—is highlighted in bold in both panels.

**Panel A.** Spline order × number of latent factors (one-layer KAN).

| Order | Sparse [2] | | Tight [4] | | Moderate [8] | | Wide [16] | |
|---|---|---|---|---|---|---|---|---|
| | VW | EW | VW | EW | VW | EW | VW | EW |
| 1 | 1.34 | 1.50 | 1.27 | 1.49 | 1.17 | 1.55 | 1.18 | 1.65 |
| 2 | 1.18 | 1.41 | 1.22 | 1.45 | 1.12 | 1.48 | 1.18 | 1.68 |
| 3 | 1.35 | 1.46 | 1.32 | 1.49 | 1.38 | 1.68 | 1.17 | 1.65 |

**Panel B.** Network depth and width (order-3 splines).

| | Deep [32,16,8] | | Modest [16,8] | | Shallow [8] | | Modest [8,4] | | Deep [8,4,2] | |
|---|---|---|---|---|---|---|---|---|---|---|
| | VW | EW | VW | EW | VW | EW | VW | EW | VW | EW |
| SR | -0.28 | -0.24 | 1.33 | 1.58 | 1.38 | 1.68 | 1.34 | 1.61 | -0.15 | 0.05 |

## Table IA2: Performance of the KAPM8 Portfolio against the Literature

This table compares the performance of the KAPM8 portfolio with the best-performing neural-network (NN4) portfolio reported by Gu, Kelly, and Xiu (2020). The full sample spans 1987–2023; the subsample spans 1987–2016 to benchmark against Gu et al. (2020). VW and EW denote value- and equally weighted high-minus-low portfolio returns. Alpha and the information ratio are computed from residual returns of the Fama–French five-factor model augmented with momentum (FF5F+MOM). Entries marked "–" were not reported by Gu et al. (2020).

| | KAPM — Our Paper | | | | NN4 — Gu et al. (2020) | |
|---|---|---|---|---|---|---|
| Data Period | 1987–2023 | | 1987–2016 | | 1987–2016 | |
| Data Sample | Excl. MicroCap | | | | Excl. MicroCap | All |
| Statistic | VW | EW | VW | EW | EW | VW |
| **Return and Risk** | | | | | | |
| Sharpe Ratio | 1.38 | 1.68 | 1.50 | 1.98 | 1.69 | 1.35 |
| Return (%) | 20.19 | 22.13 | 24.42 | 26.36 | 27.12 | 27.12 |
| Volatility (%) | 13.50 | 12.21 | 14.70 | 12.02 | 16.04 | 20.09 |
| MaxDD (%) | -30.35 | -27.07 | -26.93 | -18.31 | – | -51.78 |
| Skewness | 0.29 | 0.02 | 0.41 | 0.03 | – | – |
| Kurtosis | 1.49 | 1.57 | 1.58 | 2.13 | – | – |
| **FF5F + Momentum Test** | | | | | | |
| $\alpha$ | 1.74 | 1.78 | 1.68 | 1.91 | – | 1.76 |
| $t(\alpha)$ | 6.01 | 6.24 | 6.46 | 8.15 | – | 6.00 |
| Info Ratio | 1.31 | 1.53 | 1.34 | 1.68 | – | 1.18 |

## Table IA3: Cross-Sectional Tests — Statistical Factor Models with KAPM3 Factors

This table reports stock-level cross-sectional asset-pricing tests of the three leading KAPM3 synthetic factors on four statistical factor models estimated with K=3 components: PCA, RP-PCA, Kernel PCA, and IPCA. Panel A uses the statistical factors alone; Panel B adds the three leading KAPM3 synthetic factors ($F_1$, $F_2$, $F_3$). Fama–MacBeth *t*-statistics are reported in parentheses. The standard error of a coefficient is the time-series standard deviation of its monthly estimates divided by the square root of the number of months. *, **, *** denote significance at the 10%, 5%, and 1% levels.

| | Panel A: Without KAPM3 | | | | Panel B: With KAPM3 | | | |
|---|---|---|---|---|---|---|---|---|
| | PCA | RP-PCA | K-PCA | IPCA | PCA | RP-PCA | K-PCA | IPCA |
| $F_1$ | | | | | 1.650*** | 1.694*** | 1.704*** | 1.655*** |
| | | | | | (2.85) | (3.01) | (2.96) | (2.99) |
| $F_2$ | | | | | 2.778* | 2.763* | 2.743** | 2.845* |
| | | | | | (1.85) | (1.86) | (1.99) | (1.88) |
| $F_3$ | | | | | -0.749* | -0.745* | -0.780* | -0.774* |
| | | | | | (-1.75) | (-1.75) | (-1.74) | (-1.72) |
| PCA1 | 0.291 | | | | 0.096 | | | |
| | (0.51) | | | | (0.17) | | | |
| PCA2 | 0.069 | | | | 0.071 | | | |
| | (0.19) | | | | (0.20) | | | |
| PCA3 | -0.098 | | | | -0.092 | | | |
| | (-0.46) | | | | (-0.44) | | | |
| RP1 | | 0.138 | | | | 0.017 | | |
| | | (0.46) | | | | (0.06) | | |
| RP2 | | 0.158 | | | | 0.018 | | |
| | | (0.29) | | | | (0.03) | | |
| RP3 | | 0.126 | | | | 0.109 | | |
| | | (0.34) | | | | (0.30) | | |
| K1 | | | 2.360 | | | | 1.442 | |
| | | | (1.18) | | | | (0.74) | |
| K2 | | | 0.020 | | | | -0.261 | |
| | | | (0.01) | | | | (-0.11) | |
| K3 | | | 2.413** | | | | 2.623** | |
| | | | (2.23) | | | | (2.42) | |
| I1 | | | | -0.195 | | | | -0.140 |
| | | | | (-0.59) | | | | (-0.43) |
| I2 | | | | -0.046 | | | | 0.013 |
| | | | | (-0.23) | | | | (0.07) |
| I3 | | | | -0.095 | | | | -0.121 |
| | | | | (-0.45) | | | | (-0.58) |
| α (%) | 0.625** | 0.595** | 0.889*** | 0.650** | 0.236 | 0.181 | 0.455 | 0.316 |
| | (2.54) | (2.21) | (3.71) | (2.50) | (0.35) | (0.27) | (0.68) | (0.50) |
| $R^2$ | 0.004 | 0.003 | 0.031 | 0.006 | 0.054 | 0.053 | 0.080 | 0.058 |
| Adj. $R^2$ | 0.003 | 0.002 | 0.030 | 0.005 | 0.053 | 0.052 | 0.079 | 0.056 |
| Months | 443 | 443 | 443 | 443 | 443 | 443 | 443 | 443 |
| Stocks | 3910 | 3910 | 3910 | 3910 | 3910 | 3910 | 3910 | 3910 |

## Table IA4: Cross-Sectional Tests — Established Factor Models with KAPM3 Factors

This table reports stock-level cross-sectional asset-pricing tests of the three leading KAPM3 synthetic factors on five established factor models, each augmented with the Fama-French three factors (MKT, SMB, HML): the Hou-Xue-Zhang q-factor model, Pastor-Stambaugh liquidity, Quality-Minus-Junk, Betting-Against-Beta, and the Stambaugh-Yuan mispricing factors. Panel A excludes the KAPM3 factors; Panel B adds the three leading KAPM3 synthetic factors ($F_1$, $F_2$, $F_3$). Fama–MacBeth *t*-statistics are reported in parentheses. The standard error of a coefficient is the time-series standard deviation of its monthly estimates divided by the square root of the number of months. *, **, *** denote significance at the 10%, 5%, and 1% levels.

| | Panel A: Without KAPM3 | | | | | Panel B: With KAPM3 | | | | |
|---|---|---|---|---|---|---|---|---|---|---|
| | q-factor | Liquidity | Quality | BAB | Mispricing | q-factor | Liquidity | Quality | BAB | Mispricing |
| MKT | 0.033 | 0.002 | 0.092 | 0.023 | 0.069 | 0.103 | 0.076 | 0.163 | 0.069 | 0.137 |
| | (0.15) | (0.01) | (0.36) | (0.10) | (0.34) | (0.45) | (0.29) | (0.64) | (0.30) | (0.66) |
| SMB | -0.160 | -0.207 | -0.196 | -0.119 | -0.179 | -0.116 | -0.148 | -0.149 | -0.070 | -0.141 |
| | (-0.87) | (-1.04) | (-1.03) | (-0.63) | (-1.02) | (-0.64) | (-0.76) | (-0.78) | (-0.38) | (-0.80) |
| HML | -0.193 | -0.127 | -0.180 | -0.003 | -0.238* | -0.231* | -0.191 | -0.213 | -0.079 | -0.269** |
| | (-1.49) | (-0.82) | (-1.29) | (-0.02) | (-1.96) | (-1.77) | (-1.23) | (-1.51) | (-0.52) | (-2.20) |
| $F_1$ | | | | | | 1.767*** | 1.747*** | 1.690*** | 1.736*** | 1.738*** |
| | | | | | | (3.04) | (3.04) | (2.90) | (2.98) | (3.01) |
| $F_2$ | | | | | | 2.698* | 2.755* | 2.576* | 2.747* | 2.643* |
| | | | | | | (1.81) | (1.85) | (1.72) | (1.82) | (1.76) |
| $F_3$ | | | | | | -0.757* | -0.709* | -0.744* | -0.768* | -0.664* |
| | | | | | | (-1.90) | (-1.79) | (-1.83) | (-1.70) | (-1.93) |
| $I/A^q$ | -0.031 | | | | | -0.057 | | | | |
| | (-0.35) | | | | | (-0.64) | | | | |
| $ROE^q$ | 0.153 | | | | | 0.127 | | | | |
| | (1.59) | | | | | (1.33) | | | | |
| $EG^q$ | 0.092 | | | | | 0.061 | | | | |
| | (0.92) | | | | | (0.62) | | | | |
| LIQ | | 0.009 | | | | | 0.038 | | | |
| | | (0.05) | | | | | (0.20) | | | |
| QMJ | | | 0.218* | | | | | 0.161 | | |
| | | | (1.68) | | | | | (1.26) | | |
| BAB | | | | 0.026 | | | | | -0.037 | |
| | | | | (0.17) | | | | | (-0.25) | |
| MGMT | | | | | -0.120 | | | | | -0.173 |
| | | | | | (-0.78) | | | | | (-1.14) |
| PERF | | | | | 0.457*** | | | | | 0.402** |
| | | | | | (2.78) | | | | | (2.50) |
| α (%) | 0.592*** | 0.641** | 0.566** | 0.523*** | 0.599*** | 0.107 | 0.179 | 0.094 | 0.109 | 0.115 |
| | (2.62) | (2.51) | (2.30) | (2.84) | (3.35) | (0.16) | (0.26) | (0.14) | (0.16) | (0.17) |
| $R^2$ | 0.017 | 0.012 | 0.023 | 0.003 | 0.034 | 0.070 | 0.064 | 0.071 | 0.055 | 0.084 |
| Adj. $R^2$ | 0.016 | 0.011 | 0.022 | 0.002 | 0.033 | 0.068 | 0.063 | 0.070 | 0.053 | 0.082 |
| Months | 443 | 443 | 443 | 443 | 443 | 443 | 443 | 443 | 443 | 443 |
| Stocks | 3910 | 3910 | 3910 | 3910 | 3910 | 3910 | 3910 | 3910 | 3910 | 3910 |

## Table IA5: Cross-Sectional Tests: Robustness with Newey-West Standard Errors

This table replicates the specifications in Table 2, columns (5) and (6), and Table 6, columns (5) and (6), using Newey–West standard errors. In those tables, the standard error of a coefficient is the time-series standard deviation of its monthly estimates divided by the square root of the number of months, as in Adrian, Etula, and Muir (2014) and Borri, Chetverikov, Liu, and Tsyvinski (2024). Here, we replace the variance of the monthly estimates with a Newey–West long-run variance estimated with five lags (L=(443)^(1/4), rounded to 5). The main conclusions drawn from Table 2 and Table 6 remain unchanged. *, **, *** denote significance at the 10%, 5%, and 1% levels.

| | (1)<br>FF6 (No KAPM) | (2)<br>FF6 | (3)<br>FF6 | (4)<br>FF6 |
|---|---|---|---|---|
| KAPM8 | | 1.161*** | | |
| | | (4.04) | | |
| $F_1$ | | | 1.681*** | |
| | | | (2.61) | |
| $F_2$ | | | 2.712* | |
| | | | (1.93) | |
| $F_3$ | | | -0.706 | |
| | | | (-1.29) | |
| KAPM3 | | | | 1.150*** |
| | | | | (2.63) |
| MKT−RF | 0.016 | 0.101 | 0.063 | 0.040 |
| | (0.11) | (0.66) | (0.40) | (0.10) |
| SMB | -0.140 | -0.045 | -0.093 | -0.133 |
| | (-0.81) | (-0.27) | (-0.56) | (-0.48) |
| HML | -0.013 | -0.086 | -0.036 | -0.031 |
| | (-0.08) | (-0.55) | (-0.22) | (-0.20) |
| RMW | 0.086 | 0.003 | 0.032 | 0.065 |
| | (0.75) | (0.03) | (0.30) | (0.60) |
| CMA | -0.037 | -0.072 | -0.039 | -0.043 |
| | (-0.43) | (-0.88) | (-0.45) | (-0.42) |
| MOM | 0.081 | 0.112 | 0.088 | 0.086 |
| | (0.64) | (0.91) | (0.70) | (0.53) |
| α | 0.544*** | -0.273* | 0.141 | 0.048 |
| | (2.96) | (-1.71) | (0.26) | (0.15) |
| Adj. $R^2$ | 0.005 | 0.083 | 0.054 | 0.028 |
| $R^2$ | 0.007 | 0.085 | 0.056 | 0.030 |
| Months | 443 | 443 | 443 | 443 |
| Stocks | 3910 | 3910 | 3910 | 3910 |

## Table IA6: Polynomial Approximations of the Recovered Synthetic-Factor Functions across Support Ranges

This table reports global third-order polynomial approximations to the recovered outer-layer transformations of the three leading KAPM3 factors (plotted in Figure IA2), fit over progressively narrower central support ranges. The exercise serves as a robustness check: it shows that the cubic approximation is stable as the evaluation window shrinks from the broad 0.5–99.5 percentile interval toward the center of the distribution. Panels A, B, and C correspond to KAPM3 factors 1, 2, and 3, respectively. All approximations exceed 99.5% reconstruction accuracy.

**Panel A. Factor 1.**

| Support Range | Polynomial Approximation |
|---|---|
| 0.5–99.5% | $y = 0.01314x^3 - 0.02864x^2 - 0.004742x + 0.01747$ |
| 1–99% | $y = 0.01309x^3 - 0.02848x^2 - 0.004864x + 0.01749$ |
| 2.5–97.5% | $y = 0.01551x^3 - 0.03182x^2 - 0.003558x + 0.01736$ |
| 5–95% | $y = 0.01838x^3 - 0.03585x^2 - 0.00186x + 0.01715$ |

**Panel B. Factor 2.**

| Support Range | Polynomial Approximation |
|---|---|
| 0.5–99.5% | $y = 0.05159x^3 + 0.05454x^2 - 0.01316x - 0.01249$ |
| 1–99% | $y = 0.04207x^3 + 0.03874x^2 - 0.02061x - 0.01337$ |
| 2.5–97.5% | $y = 0.01783x^3 - 0.004323x^2 - 0.04341x - 0.01675$ |
| 5–95% | $y = -0.008161x^3 - 0.05508x^2 - 0.07432x - 0.02251$ |

**Panel C. Factor 3.**

| Support Range | Polynomial Approximation |
|---|---|
| 0.5–99.5% | $y = 0.02832x^3 + 0.09171x^2 + 0.08488x + 0.02021$ |
| 1–99% | $y = 0.03457x^3 + 0.1059x^2 + 0.0954x + 0.02273$ |
| 2.5–97.5% | $y = 0.04188x^3 + 0.1236x^2 + 0.1093x + 0.02629$ |
| 5–95% | $y = 0.04302x^3 + 0.1282x^2 + 0.1141x + 0.02778$ |

## Figure IA1: In-Sample Performance of KAPM Portfolios by First- and Second-layer Architectures

Panel A reports the annualized in-sample Sharpe ratio of KAPM long–short portfolios under different architectures, for the smallest training sample (1957-1986) on the left and largest training sample (1957-2022) on the right. The Sharpe ratios are produced by high-minus-low portfolios, which are produced from the return predictions of the candidate architectures. The label "none" on the horizontal axis means that there is no second layer. We present the results in a blue gradient coloring where darker values indicate larger Sharpe ratio values.

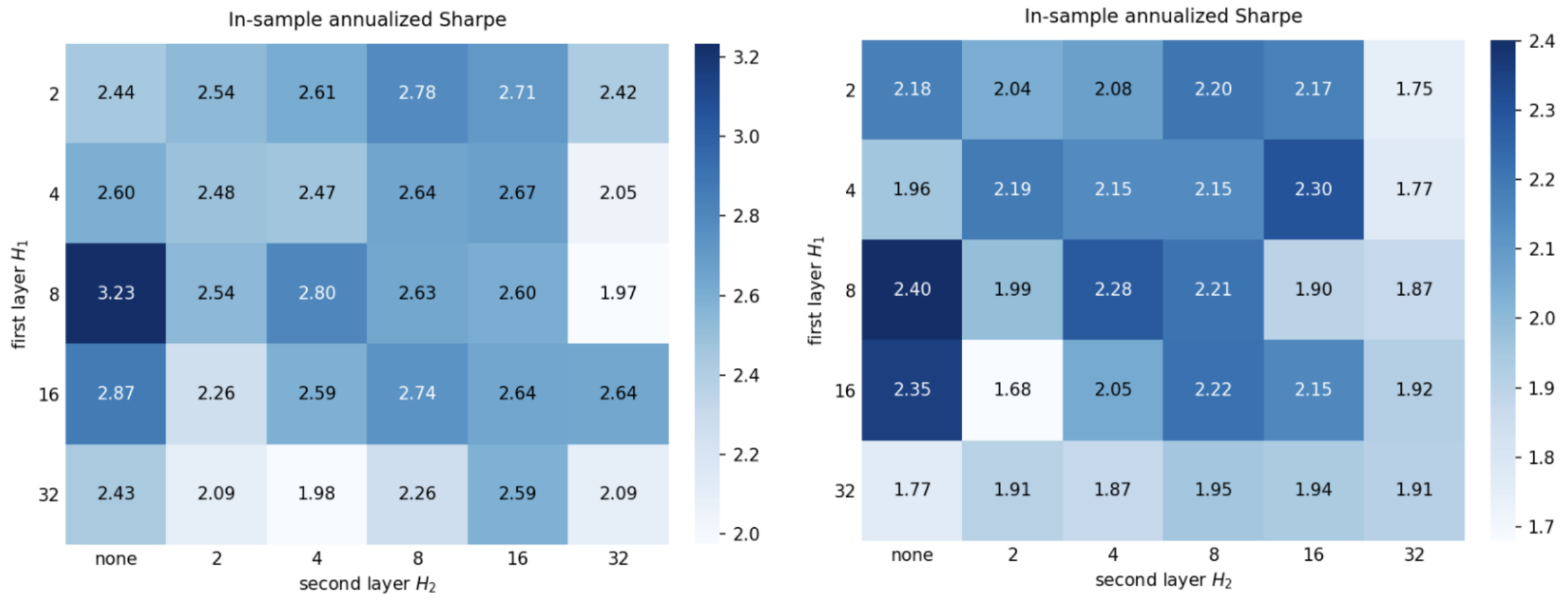

Panel B reports the inclusion probabilities of each in-sample high-minus-low portfolio return series, when compared against all other 24 series. The return series run through the smallest training sample (1957-1986) on the left and largest training sample (1957-2022) on the right. The label "none" on the horizontal axis means that there is no second layer. A high inclusion probability indicates that the portfolio is not spanned by the portfolios of the other architectures; a low inclusion probability indicates that its information is already captured by the others.

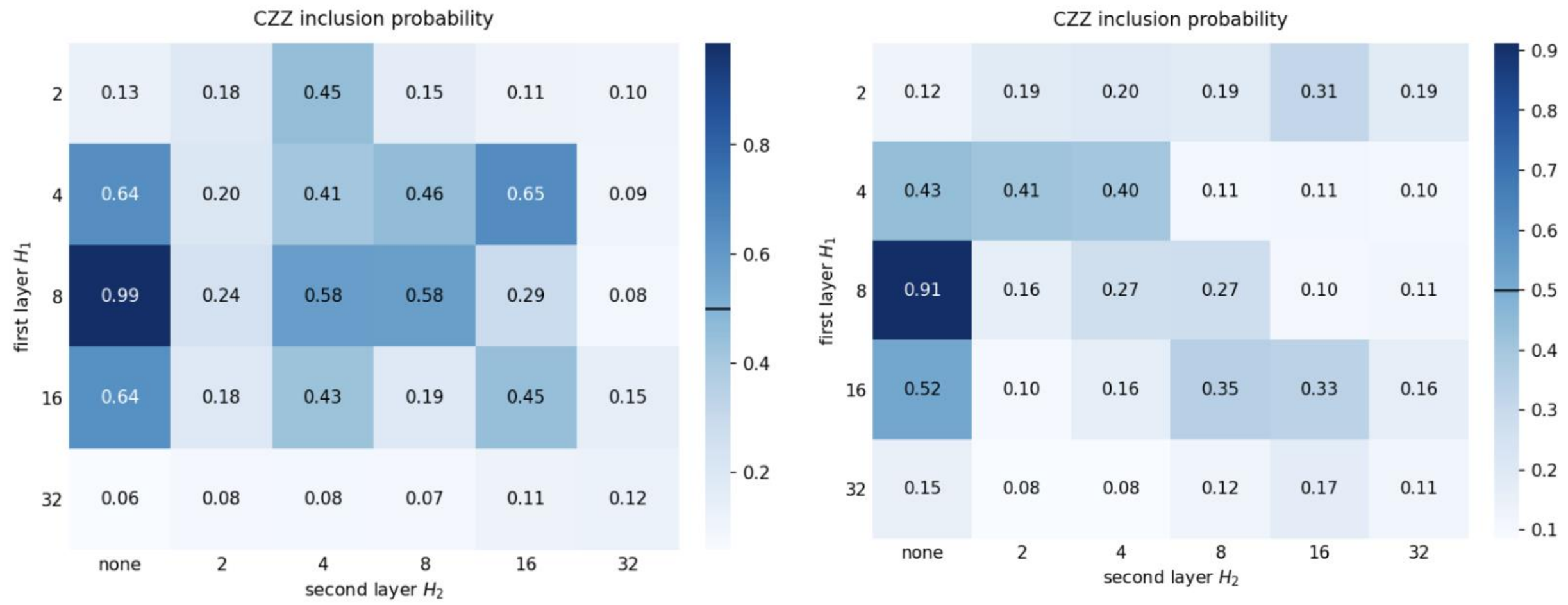

## Figure IA2: Approximate Equations and Visualization of KAPM3 Factors

This figure plots the recovered outer-layer transformations $\Phi_q(\cdot)$ of the KAPM3 factors (panels (a)–(c)), each evaluated over the 0.5th–99.5th percentile range of its hidden-node activation. The globally accurate cubic approximations of the recovered spline functions are:

$$\widetilde{\Phi}_1(x) = 0.01747 - 0.004742\,x - 0.02864\,x^2 + 0.01314\,x^3$$

$$\widetilde{\Phi}_2(x) = -0.01249 - 0.01316\,x + 0.05454\,x^2 + 0.05159\,x^3$$

$$\widetilde{\Phi}_3(x) = 0.02021 + 0.08488\,x + 0.09171\,x^2 + 0.02832\,x^3$$

Internet Appendix Section 5.1 describes the procedure that recovers the spline functions and produces a single accurate cubic approximation; the robustness of these approximations across narrower support ranges is documented in Internet Appendix Table IA6. Visually, the first function is approximately linear and decreasing over its support; the second is predominantly decreasing, with a flatter left tail and modest curvature in the right tail; and the third exhibits a pronounced convex (U-shaped) profile.

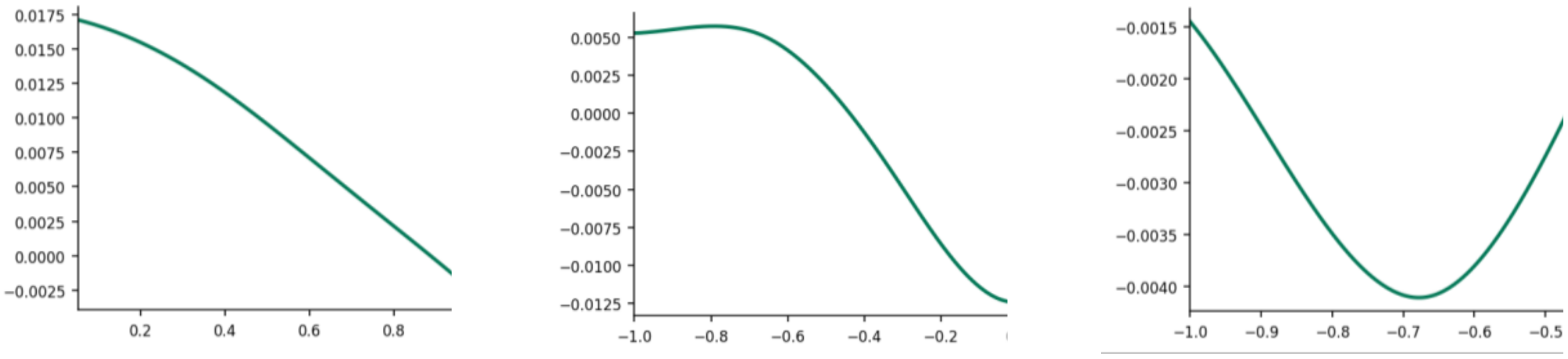

**(a) Factor 1** **(b) Factor 2** **(c) Factor 3**